\documentclass[twocolumn]{aastex61}
\usepackage{amsmath}
\usepackage{color}
\usepackage{enumitem}
\setlist{nolistsep}

\usepackage{ulem}

\revised{}
\submitjournal{ApJ}

\shorttitle{Tests of the TW Method}
\shortauthors{Zou et al.}

\begin{document}

\title{Testing the Tremaine-Weinberg Method Applied to Integral-field
  Spectroscopic Data Using a Simulated Barred Galaxy}



\author[0000-0002-0786-7307]{Yanfei Zou}
 \affiliation{SOA Key Laboratory for Polar Science, Polar Research 
 Institute of China, 451 Jinqiao Road, Shanghai, 200136, China; zouyanfei@pric.org.cn}
 \affiliation{Key Laboratory of Research in Galaxies and Cosmology,
  Shanghai Astronomical Observatory, Chinese Academy of Sciences, \\
  80 Nandan Road, Shanghai 200030, China}
\affiliation{School of Astronomy and Space Sciences, University of
  Chinese Academy of Sciences, 19A Yuquan Road, Beijing 100049, China}

\author[0000-0001-5604-1643]{Juntai Shen} 
\affiliation{Department of Astronomy, School of Physics and Astronomy,
  Shanghai Jiao Tong University, 800 Dongchuan Road, Shanghai 200240, China; jtshen@sjtu.edu.cn; lizy.astro@sjtu.edu.cn}
\affiliation{Key Laboratory of Research in
  Galaxies and Cosmology,
  Shanghai Astronomical Observatory, Chinese Academy of Sciences, \\
  80 Nandan Road, Shanghai 200030, China}
\affiliation{School of Astronomy and Space Sciences, University of
  Chinese Academy of Sciences, 19A Yuquan Road, Beijing 100049, China}

\author[0000-0003-4980-1012]{Martin Bureau}
\affiliation{Sub-department of Astrophysics, Department of Physics,
  University of Oxford, Denys Wilkinson Building, Keble Road, \\
  Oxford OX1 3RH, UK; martin.bureau@physics.ox.ac.uk}
\affiliation{Yonsei Frontier Lab and Department of Astronomy, Yonsei
  University, 50 Yonsei-ro, Seodaemon-gu, Seoul 03722, Republic of
  Korea}

\author[0000-0001-5017-7021]{Zhao-Yu Li}
\affiliation{Department of Astronomy, School of Physics and Astronomy,
  Shanghai Jiao Tong University, 800 Dongchuan Road, Shanghai 200240, China; jtshen@sjtu.edu.cn; lizy.astro@sjtu.edu.cn}
\affiliation{Key Laboratory of Research in Galaxies and Cosmology,
  Shanghai Astronomical Observatory, Chinese Academy of Sciences, \\
  80 Nandan Road, Shanghai 200030, China}


\begin{abstract}
Tremaine and Weinberg (TW) proposed a conceptually simple procedure
relying on long-slit spectroscopy to measure the pattern speeds of
bars ($\Omega_{\rm p}$) in disk galaxies. Using a simulated galaxy, we
investigate the potential biases and uncertainties of TW measurements
using increasingly popular integral-field spectrographs (IFSs), for
which multiple pseudo-slits (and thus independent measurements) can be
constructed with a single observation. Most importantly, to establish
the spatial coverage required and ensure the validity of the
measurements, the inferred $\Omega_{\rm p}$ must asymptotically
converge as the (half-)length of each pseudo-slit used is
increased. The requirement for our simulation is to reach $\approx1.3$
times the half-light radius, but this may vary from galaxy to
galaxy. Only those slits located within the bar region yield accurate
measurements. We confirm that the position angle of the
  disk is the dominant source of systematic error in TW
  $\Omega_{\rm p}$ measurements, leading to under/overestimates of
  tens of percent for inaccuracies of even a few degrees. Recasting
the data so that the data grid aligns with the disk major axis leads
to slightly reduced uncertainties. Accurate measurements are obtained
only for well-defined ranges of the bar angle (relative to the galaxy
major axis) $\phi_{\rm bar}$ and the inclination angle $i$,
here $10\degr\la\phi_{\rm bar}\la75\degr$ and
  $105\degr\la\phi_{\rm bar}\la170\degr$ and $15\degr\la i\la70\degr$. The
adopted (pseudo-)slit widths, spatial resolution, and (unless
extremely aggressive) spatial binning of IFS data have no significant
impact on the measurements. Our results thus provide useful guidelines
for reliable and accurate direct $\Omega_{\rm p}$ measurements with
IFS observations.
\end{abstract}


 
\section{Introduction}
\label{sec:intro}

Bars are present in at least half of nearby disk galaxies
\citep[e.g.][]{esk2000,mar2007,men2007,bar2008,agu2009}. They play an
important role in the redistribution of angular momentum and energy
across the different components of the host disk, as well as between
the disk and dark matter halo
\citep[e.g.][]{wei1985,deb1998,deb2000,ath2003}. The dynamics of
barred galaxies depends primarily (but not exclusively) on the angular
velocity or pattern speed of the bar, denoted $\Omega_{\rm
  p}$. Usually, $\Omega_{\rm p}$ is parameterized by the dimensionless
bar rotation rate $\mathcal{R}\equiv R_{\rm CR}/a_{\rm bar}$, where
$a_{\rm bar}$ is the half-length of the bar and the corotation radius
$R_{\rm CR}$ (where the bar rotation speed equals that of disk
material) can be determined from the disk circular velocity curve and
$\Omega_{\rm p}$. Fast bars are defined as having
$1\leq\mathcal{R}\leq1.4$ while slow bars have $\mathcal{R}>1.4$
\citep{deb2000}; $\mathcal{R}<1$ is not allowed by the orbital
structure of barred disks \citep[e.g.][]{con1980,ath1992a}.

The measurement of $\Omega_{\rm p}$ is challenging. There are a
variety of methods in the literature to indirectly measure bar pattern
speeds, which are all model dependent. In particular, hydrodynamical
simulations of individual barred galaxies have been used to infer
$\Omega_{\rm p}$, by matching the observed gas morphology and/or gas
velocities with those of a simulated galaxy
\citep[e.g.][]{wei2001,rau2008,tre2008,frag2017}. There are also some
physically motivated methods to measure $\Omega_{\rm p}$, associating
particular morphological features with Lindblad
resonances. Specifically, these methods associate the position of
galaxy rings with certain resonances
\citep[e.g.][]{buta1986,buta1995,veg1997,muo2004,per2012}, they
examine the offsets and shapes of dust lanes
\citep[e.g.][]{van1982,ath1992b}, they analyze the changes of the
morphologies and/or phases of spiral arms near $R_{\rm CR}$
\citep[e.g.][]{can1993,can1997,pue1997,agu1998,buta2009}, they
identify color and star formation changes beyond the bar region
\citep[e.g.][]{cep1990,agu2000}, and they characterize the gas and/or
stellar velocity residuals after subtraction of the (axisymmetric)
rotation pattern
\citep[e.g.][]{sem1995b,font2011,font2014,bec2018}. Nevertheless, to
gauge their accuracies, most of these indirect methods must still be
compared with direct measurements.

The only model-independent way to measure $\Omega_{\rm p}$ is the
Tremaine-Weinberg (TW) method \citep{tre1984}, which uses a simple and
elegant formalism to infer $\Omega_{\rm p}$ directly from the data. In
this paper, we use an $N$-body simulation to understand the limitations,
biases, and uncertainties that can affect TW measurements,
particularly when carried out using integral-field spectrographs
(IFSs), which necessarily have limited fields of view (FOVs) and
angular resolutions (point spread functions [PSFs]). We first review in
\S~\ref{sec:tw} the TW formalism and its likely limitations, before
describing in \S~\ref{sec:model} the simulation and mock data sets used
to quantify those limitations. \S~\ref{sec:results} presents analyses
of a number of specific potential biases and discusses their likely
impact on real measurements. \S~\ref{sec:concl} briefly summarizes the
main results.

\section{Tremaine-Weinberg Method}
\label{sec:tw}

\subsection{Formalism}
\label{subsec:tw-formalism}

As mentioned above, the TW method is the only model-independent means
to infer $\Omega_{\rm p}$ from observational data. Its original and
simplest implementation uses long-slit spectroscopic observations,
under the assumptions that the disk is flat and the bar (actually the
whole disk) has a single well-defined pattern speed. Another key
requirement is that the (surface brightness of the) tracer population
obeys the continuity equation. From these assumptions,
$\Omega_{\rm p}$ can be derived as
\begin{subequations}
  \label{eq:tw-method}
  \begin{align}
\Omega_{\rm p}\sin i &\,=\,\frac{\int_{-\infty}^{+\infty} h(Y)\,dY\int_{-\infty}^{+\infty}\,[V_{||}(X,Y)\!\!-\!\!V_{\rm sys}]\,\Sigma(X,Y)\,dX}{\int_{-\infty}^{+\infty} h(Y)\,dY\int_{-\infty}^{+\infty}\,[X\!\!-\!\!X_{\rm c}]\,\Sigma(X,Y)\,dX}{}\label{eq:tw} \\
                    & {}\,=\,\frac{<V_{||}>\!-V_{\rm sys}}{<X>\!-X_{\rm c}}\,\,\,,\label{eq:mk}
  \end{align}
\end{subequations}
where ($X$,$Y$) are sky plane coordinates with origin at the galaxy
center, $V_{||}(X,Y)$ is the mean line-of-sight (LOS) velocity and
$\Sigma(X,Y)$ the surface brightness (or surface mass density) of the
tracer adopted, $V_{\rm sys}$ is the galaxy systemic velocity,
$X_{\rm c}$ is the position of the galaxy minor axis, and $h(Y)$ is an
arbitrary weighting function. The $X$-axis and the $Y$-axis must be aligned
with the major and the minor axis of the galaxy disk, respectively, and
the slits (one for each $Y$) should thus be aligned parallel to but
offset from the disk major axis (so that the integrals along $X$
correspond to summations along slits parallel to but offset from the
major axis).

\cite{mer1995} refined the TW method, suggesting to simply collapse
the long-slit spectra to yield higher signal-to-noise ratio (S/N)
measurements. They thus rewrote Equation~\ref{eq:tw-method}(a) into the
equivalent Equation~\ref{eq:tw-method}(b), where
\begin{equation}
  \label{eq:mk-v}
<V_{||}>\,\equiv\,\frac{\int_{-\infty}^{+\infty}\,V_{||}(X,Y)\,\Sigma(X,Y)\,dX}{\int_{-\infty}^{+\infty}\,\Sigma(X,Y)\,dX}
\end{equation}
and
\begin{equation}
  \label{eq:mk-x}
<X>\,\equiv\,\frac{\int_{-\infty}^{+\infty}\,X\,\Sigma(X,Y)\,dX}{\int_{-\infty}^{+\infty}\,\Sigma(X,Y)\,dX}
\end{equation}
are, respectively, the luminosity-weighted mean velocity and the mean
position obtained from collapsing one long-slit spectrum along,
respectively, its spatial and its dispersion (wavelength)
direction. Then, $\Omega_{\rm p}\sin i$ can easily be obtained from the
slope of a straight line fit to $<V_{||}>$ versus $<X>$ for multiple
slits. In addition to a single higher-S/N measurement for each of
the numerator and denominator in Equation~\ref{eq:tw-method}(b), another
advantage of this procedure is that the uncertainties on $V_{\rm sys}$
and $X_{\rm c}$ are unimportant (while they can lead to systematic
biases when using Eq.~\ref{eq:tw-method}(a) directly). We will refer to this
method as the MK method.

The TW method infers the bar pattern speed by quantifying the
asymmetries introduced by the bar in the galaxy disk. Indeed, for an
axisymmetric disk with no bar, $\Sigma(X,Y)$ is an even function of
$X$ (i.e.,\ along the slit), while $V_{||}(X,Y)$ and $X$ are odd
functions of $X$. Thus, the integrals in the numerator and denominator
of Equation~\ref{eq:tw-method} both sum to zero, and $\Omega_{\rm p}$
is undefined. The existence of a bar (or a spiral pattern) introduces
an oddness into $\Sigma(X,Y)$ and an evenness into $V_{||}(X,Y)$,
resulting in nonzero integrals and the desired measurement of
$\Omega_{\rm p}$. This is the main power of the TW method but also its
greatest weakness, in that the integrals in the numerator and
denominator of Equation~\ref{eq:tw-method} effectively take the
difference of two large numbers, making TW measurements extremely
sensitive to imperfections in the data and prone to systematic
biases. In particular, asymmetries due to (sub)structures other than
the bar can lead to false signals. It is thus essential to ensure that
the contributions to the integrals in Equation~\ref{eq:tw-method} are
indeed due to the bar and not other artifacts.

There are various ways to measure bar pattern speeds using the TW
method. According to Equation~\ref{eq:tw-method}(a), $\Omega_{\rm p}$ can be
measured from a single measurement using a single slit or from the
average of multiple measurements using multiple slits
\citep[e.g.][]{ken1987,bur1999}. We will refer to the latter method as
the TW averaging method. As suggested by Equation~\ref{eq:tw-method}(b), another
way to measure $\Omega_{\rm p}$ is to measure the slope of $<V_{||}>$
versus $<X>$. However, instead of collapsing the spectra to measure
$<V_{||}>$ and $<X>$, as suggested by \cite{mer1995}, $<V_{||}>$ and
$<X>$ can equally be derived from measurements of $\Sigma(X,Y)$ and
$V_{||}$ at each position along the slits, following
Equations~\ref{eq:mk-v} and \ref{eq:mk-x} \citep[see,
e.g.,][]{agu2015}. We will refer to this as the TW fitting
method. Overall, there are thus at least four variants of the TW
method: the original TW method, the TW averaging method, the TW
fitting method, and the MK method. The first two are sensitive to
uncertainties in $V_{\rm sys}$ and $X_{\rm c}$ while the last two are
not.

The most important assumption of the TW method is arguably that the
tracer population satisfies the continuity equation. Clearly, the old
stellar populations of S0 galaxies are good targets for the TW method,
as star formation (SF) and dust extinction are usually unimportant.
As a result, S0 galaxies have been frequent targets of TW studies
\citep[e.g.][]{ken1987,mer1995,ger1999,deb2002,agu2003,cor2003,ger2003,deb2004,cor2007}.
However, the TW method has also been applied to late-type galaxies,
despite risks that SF and dust will affect the measurements
\citep[e.g.,][]{ger2003,tre2007,agu2015,guo2019}. The application of the
TW method to gas tracers rather than stars is also more recent and
requires more caution. In galaxies, gas will normally cycle through
multiple phases (molecular, atomic, and ionized), and any given phase
may not obey the continuity equation. However, if the interstellar
medium (ISM) is dominated by a single phase, the gas in that dominant
phase may then approximately satisfy the continuity equation. TW bar
pattern speed measurements have been carried out using observations of
\ion{H}{1} \citep[e.g.,][]{bur1999,ban2013} and H$_{\rm 2}$
\citep[e.g.,][]{ran2004, zim2004}. The ionized gas rarely dominates the
ISM by mass, and any line flux does not reliably trace its mass.
However, using $N$-body/SPH simulations, \cite{her2005} found that the
TW method could be applied to the ionized gas component to get a rough
estimate of $\Omega_{\rm p}$, on the condition that the shock regions
are avoided and the measurements are confined to the gaseous bar
region. H$\alpha$ has indeed been used numerous times to measure bar
pattern speeds with the TW method
\citep[e.g.][]{her2005,ems2006,fat2007,fat2009,che2009,gab2009}. Most
of the above measurements, irrespective of the tracer, suggest that
bars rotate fast.

%
%

\subsection{Application to IFS Data}
\label{subsec:tw-ifs}

Observations with IFSs, which yield 3D data, are
becoming increasingly popular, and sizeable samples of barred galaxies
with a wide range of morphologies can now be selected from large,
homogeneous, and often publicly available IFS surveys -- e.g.\ the
Calar Alto Legacy Integral-field Spectroscopy Area Survey
\citep[CALIFA;][]{san2012}, Sydney-Australian-Astronomical-Observatory
Multi-object Integral-field Spectrograph \citep[SAMI;][]{croo2012}
survey, and Mapping Nearby Galaxies at Apache Point Observatory
\citep[MaNGA;][]{bun2015} survey. IFS observations have unique
advantages over long-slit observations for TW measurements. In
particular, multiple pseudo-slits can be constructed from a single IFS
observation. Most importantly, \citeauthor{deb2003}
  (\citeyear{deb2003}, hereafter D03) investigated slit misalignments
  and found that errors in the position angle of the disk
  (PA$_{\rm disk}$; e.g.\ from intrinsic disk ellipticities) and thus
  of the slits significantly and systematically affect TW
  $\Omega_{\rm p}$ measurements. IFSs naturally allow us to choose the
orientation of the pseudo-slits {\em after} the observations are
obtained, thus allowing us to test for systematic uncertainties
associated with the choice of PA$_{\rm disk}$.

The use of the TW method with IFS data also differs from that with
long-slit observations in several ways.

1. Extensive spatial coverage is very important to the TW method, as
the integrals in Equation~\ref{eq:tw-method}(a) formally range from $-\infty$ to
$+\infty$ \citep[see,
e.g.,][]{deb2004,ran2004,zim2004,che2009,fat2009,agu2015,guo2019}. In
practice, the integrals only need to reach the (projected) radius
where the disk becomes (nearly) axisymmetric, as the integrands will
cancel out at larger distances. There is no specific prescription to
identify this radius, and it is likely to be different
  from galaxy to galaxy. Most long slits extend to the disk
outskirts, where $\Omega_{\rm p}$ measurements should asymptotically
converge before the slits end. However, many IFSs have a small FOV,
and the disks observed may not reach axisymmetry within that
FOV. Before using the TW method with IFS data, it is thus essential to
establish the minimum spatial coverage required for reliable and
accurate TW measurements. This is done in \S~\ref{subsec:convergence}.

2. Similarly, IFS data allow us to create (i.e.,\ overlay on the
observations) an arbitrary number of pseudo-slits, at arbitrary
positions (i.e.,\ at arbitrary offsets from the galaxy major
axis). However, it is unclear whether all the resulting measurements will
be equally reliable, as the photometric and kinematic signatures of
bars necessarily decrease with increasing radius. In
\S~\ref{subsec:out_slits}, we will thus explore how the quality of a
measurement varies as a function of the slit position.

3. Spatial binning is commonly used with IFSs to increase
  the S/N of the data within a given spatial
  region (i.e.,\ a bin made up of several spaxels), especially in the
  outer parts of galaxies, where convergence of the TW integrals is
  necessary. The flux and velocity of each bin are thus altered and
  effectively replaced by their (luminosity-weighted) spatial
  averages. As these are the quantities that underlie the TW method,
  the positions and sizes of the bins are likely to impact TW
  $\Omega_{\rm p}$ measurements. In \S~\ref{subsec:spt_bin}, we will
  therefore quantify the effects of spatial binning on these
  measurements.

4. As the position angle of the disk (PA$_{\rm disk}$) is generally
not aligned with the grid/axes of the IFS data
$(X_{\rm grid},Y_{\rm grid})$, the footprints of pseudo-slits are
irregular in shape (at least in terms of the integer spaxels contained
within them), an effect absent with long-slit observations.  This
introduces extra asymmetries and can thus artificially contribute to
the integrals in Equation~\ref{eq:tw-method}(a), potentially affecting the
resulting $\Omega_{\rm p}$ measurements. A possible way to eliminate
this effect is to calculate the weights/contributions of fractional
spaxels along the slits. However, a more straightforward and practical
solution is often to simply recast the data grid. We must then test to
what extent this grid recasting can affect the measurements, and this
is done in \S~\ref{subsec:recast}.

\subsection{Application to Both IFS and Long-slit Data}
\label{subsec:tw-ifs-slit}

There are other issues that may affect TW bar pattern speed
measurements using either IFS or long-slit observations.

1. A clear advantage of using IFS over long-slit data is
  that the slits' orientation (implicitly the disk position angle
  PA$_{\rm disk}$) can readily be changed during the TW
  calculations. Although PA$_{\rm disk}$ does not appear explicitly in
  Equation~\ref{eq:tw-method}, and consequently PA$_{\rm disk}$
  uncertainties are not normally propagated through $\Omega_{\rm p}$
  uncertainties, PA$_{\rm disk}$ directly affects $\Omega_{\rm p}$
  measurements using both IFS and long-slit spectra. According to D03,
  we expect errors on PA$_{\rm disk}$ to be the largest (and
  systematic) source of error in TW $\Omega_{\rm p}$ measurements, as
  the measured $\Omega_{\rm p}$ can be under- or overestimated by
  $\approx50\%$ for a misalignment angle $\delta$PA$_{\rm disk}$ of
  only $\approx5\degr$. In particular, it might be that some of the
  so-called ultrafast bars recently reported in the literature
  \citep{guo2019} are due to such misalignments. We therefore carry
  out tests analogous to those of D03 in \S~\ref{subsec: pa_error},
  exploring a larger range of misalignment angles and different disk
  view angles. We also explore potential empirical disk position angle
  diagnostics (\S~\ref{sec: pa_diagnostic}).

2. The orientation of a barred disk with respect to the
observer is very important. In particular, the bar angle relative to
the galaxy major axis ($\phi_{\rm bar}$) is crucial. Indeed, if the
bar is exactly parallel or perpendicular to the disk major axis, the
surface brightness distribution and velocity field will remain
symmetric with respect to the disk minor axis. The integrands in
Equation~\ref{eq:tw-method}(a) will thus remain odd, again yielding an undefined
$\Omega_{\rm p}$. As TW measurements utilize $V_{||}$, nearly face-on
($i\approx0\degr$) galaxies will also yield unreliable measurements,
while bars are difficult to identify in nearly edge-on
($i\approx90\degr$) systems (the intrinsic thickness of the disks is
then also apparent and may affect the results). We will thus identify
the optimal ranges of the relative bar angle and inclination in
\S~\ref{subsec:iphi}, to derive a prescription to exclude a priori
galaxies with inappropriate orientations.

3. There is also no specific rule to choose the optimal
slit width to use. Generally, a slit width roughly equal
  to the PSF has been adopted. \cite{guo2019} tested two different
  pseudo-slit widths and reported that they both yielded consistent
  measurements. We will explicitly test the influence of the slit
  width on TW $\Omega_{\rm p}$ measurements in \S~\ref{subsec:slit}.

4. Lastly, the limited spatial resolution of the data may affect
measurements. Indeed, observations do not yield the intrinsic surface
brightness $\Sigma(X,Y)$ and mean velocity $V_{||}(X,Y)$ of the chosen
tracer, but rather those quantities convolved by the PSF of the
observations $W(X,Y)$, such that the observed surface brightness
$\Sigma_{\rm o}(X,Y)$ and velocity field $V_{\rm o}(X,Y)$ are given by
\begin{equation}
  \label{eq:seeing-sb}
    \Sigma_{\rm o}(X,Y)=\int W(X\!\!-\!\!X^{\prime},Y\!\!-\!\!Y^{\prime})\,\Sigma(X^{\prime},Y^{\prime})\,dX^{\prime}dY^{\prime}
\end{equation}    
and
\begin{equation}
  \label{eq:seeing-v}
  V_{\rm o}(X,Y)=\frac{\int W(X\!\!-\!\!X^{\prime},Y\!\!-\!\!Y^{\prime})\,\Sigma(X^{\prime},Y^{\prime})\,V_{||}(X^{\prime},Y^{\prime})\,dX^{\prime}dY^{\prime}}{\Sigma_{\rm o}(X,Y)}\,\,\,.
\end{equation}

\cite{tre1984} argued that the limited spatial resolution of
observations is unimportant if the PSF $W(X,Y)$ is
an even function of $X$, which is usually the case. To verify this,
however, we will directly test in \S~\ref{subsec:seeing} how the PSF
affects TW $\Omega_{\rm p}$ measurements.

\section{Simulation and Mock Data}
\label{sec:model}

We use a simulated barred galaxy to investigate all the issues
mentioned above (\S~\ref{subsec:tw-ifs} and
\ref{subsec:tw-ifs-slit}). We first generate mock data sets, and then
use them to test the likely influence on real TW measurements of a
number of key observational and analysis parameters (FOV,
pseudo-slit selection, spatial binning, grid recasting,
errors in disk position angle, relative bar angle and
inclination, slit width, and PSF).

\subsection{Simulation}
\label{subsec:simulation}

\begin{figure}[!t]
  \centering
  \includegraphics[width=85mm]{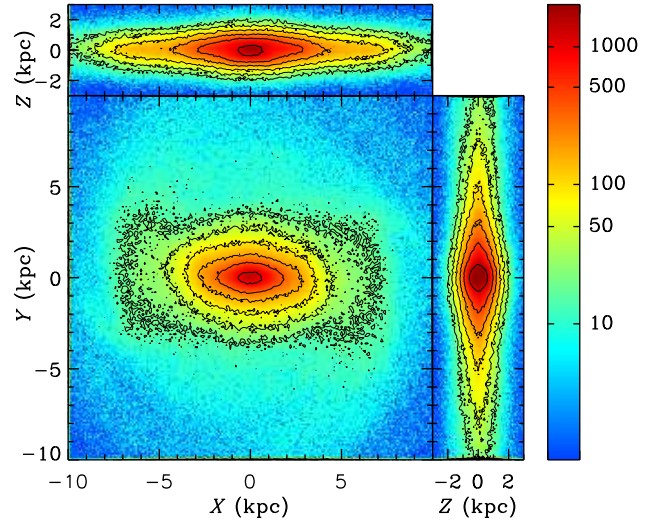}
  \caption{Face-on (center), end-on (right), and side-on (top) views of
    our simulated barred galaxy at a time of $2.4$~Gyr. Colors
    represent the surface density in arbitrary units. Black contours
    indicate $95\%$, $80\%$, $70\%$, $60\%$, $50\%$, and $40\%$ percent of the
    maximum surface density.}
  \label{fig:show_model}
\end{figure}

The $N$-body simulation we adopt is shown in
Figure~\ref{fig:show_model}. It is a simple simulation of a disk
galaxy with $10^6$ live particles of total mass
$M_{\rm d}=4.25\times10^{10}$~$M_{\odot}$, evolving in a rigid dark
matter halo spherical potential
$\Phi(r)=\frac{1}{2}V_{\rm c}^{2}\ln(1+r^{2}/r_{\rm c}^2)$, where $r$
is the spherical radius, $V_{\rm c}\approx250$~km~s$^{-1}$ is the
asymptotic circular velocity at infinity, and $r_{\rm c}=15$~kpc is
the core radius. The particles are initially distributed in a
dynamically cold (Toomre's $Q\approx1.5$) axisymmetric exponential
disk with a scale length of $\approx1.9$~kpc and a scale height
$\approx0.2$~kpc. A bar forms spontaneously and quickly buckles in the
vertical direction. To create our mock data sets, we selected a
snapshot at $2.4$~Gyr (shown in Fig.~\ref{fig:show_model}), when the
bar is quasi-steady.

We measure a maximum bar ellipticity
$\epsilon_{\rm bar, max}\approx0.47$ using the Image Reduction
  and Analysis Facility \citep[\textsc{IRAF};][]{tody1986,tody1993}
Ellipse task, from the first maximum of the radial profile of
ellipticity of the isophotes. We estimate the bar (half-)length to be
$a_{\rm bar}\approx4.6$~kpc, from the average of the radius of the
first minimum in the radial ellipticity profile (4.7~kpc) and the
radius suggested by the bar-interbar contrast method (4.5~kpc; see
\citealt{agu2000}).\footnote{The radius of the first maximum in the
  radial ellipticity profile is also often advocated as a measure of a
  bar's (half-)length. We do not consider this measure here, as it
  significantly underestimates the bar (half-)length estimated
  visually.} This maximum ellipticity and the (half-)length of the bar
compare favorably to those of typical bars \citep[see,
e.g.,][]{mar2007}, although we could of course rescale our simulation
to any desired physical size. The simulated galaxy effective
(half-mass) radius is $R_{\rm e}=2.4$~kpc. Our simulated barred galaxy
is thus well suited to realistic tests of TW measurements.

\begin{figure}[!t]
  \centering
  \includegraphics[width=80mm]{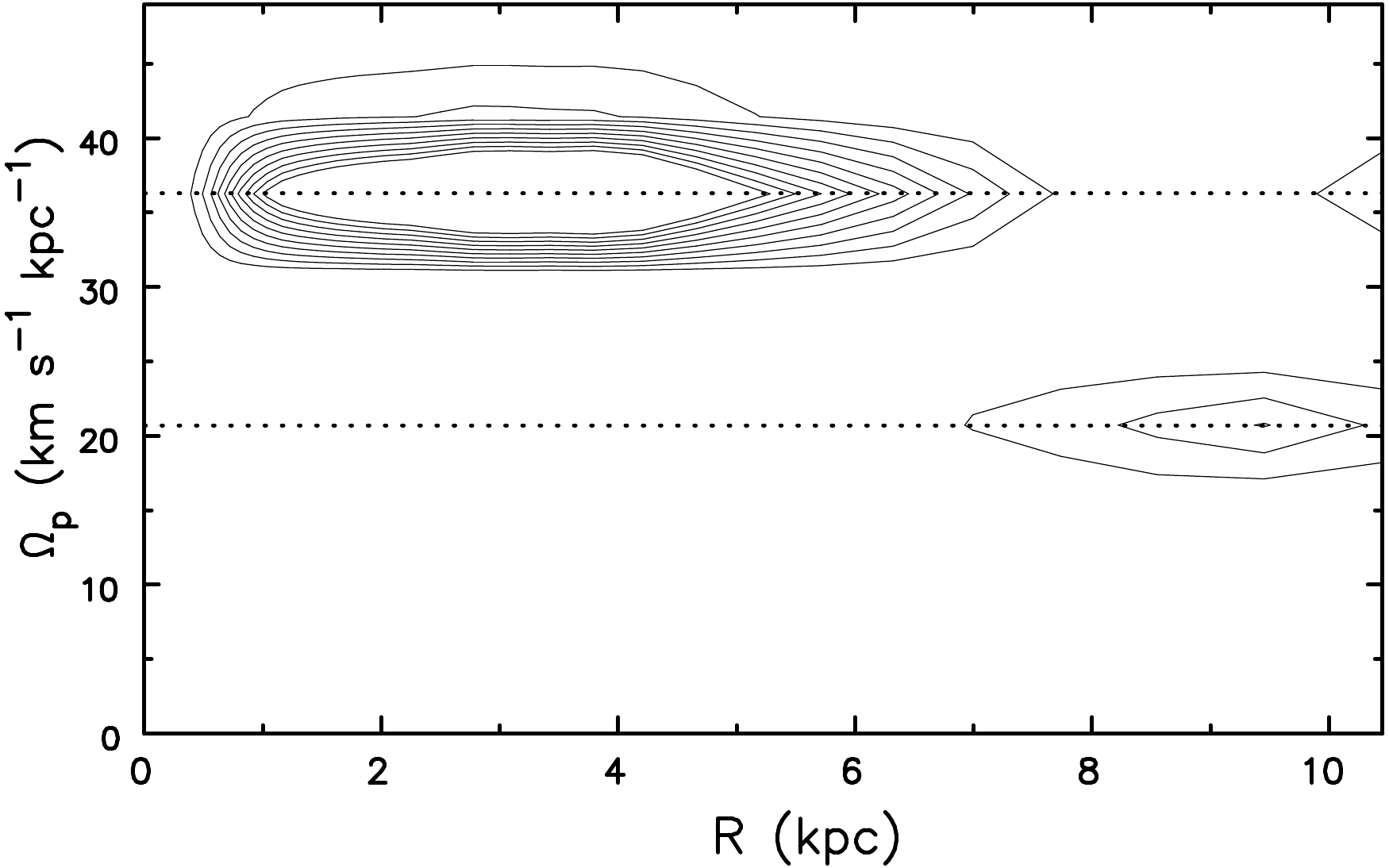}
  \caption{Contours of the $m=2$ Fourier amplitudes of density
    distribution features as a function of radius and pattern speed,
    at our selected time of $2.4$~Gyr. The pattern speeds are derived
    from a Fourier analysis of the particles at a given radius as a
    function of time. The two dotted lines along the two prominent
    ridges represent the intrinsic pattern speeds of the bar
    ($\Omega_{\rm p,int}=36.3$~km~s$^{-1}$~kpc$^{-1}$) and spiral arms
    ($\Omega_{\rm p,spiral}=20.7$~km~s$^{-1}$~kpc$^{-1}$). For
    clarity, contours are from $5\%$ to $50\%$ of the maximum only, in
    steps of $5\%$. As expected, the strength of the bar feature
    decreases rapidly outside of the adopted bar (half-)length.}
  \label{fig:radial_ps}
\end{figure}

As a function of radius, Figure~\ref{fig:radial_ps} shows the $m=2$
Fourier amplitudes of features in the density distribution that have
well-defined frequencies, thus revealing coherent patterns and their
pattern speeds (measured using several snapshots closely spaced
temporarily around our selected time of $2.4$~Gyr; see
\citealt{spa1987} for more details). The figure clearly shows that the
bar has a constant intrinsic pattern speed
$\Omega_{\rm p,int}=36.3$~km~s$^{-1}$~kpc$^{-1}$ throughout its
extent, while a pair of transient spiral arms with a lower pattern
speed ($\Omega_{\rm p,spiral}=20.7$~km~s$^{-1}$~kpc$^{-1}$) dominate
the disk beyond the bar region.


\subsection{Mock data sets}
\label{subsec:mocks}

At first, we align the simulated bar with the data $X_{\rm grid}$-axis, 
as shown in Figure~\ref{fig:show_model}. To create multiple mock
data sets, the simulation is then rotated counterclockwise around the
$Z_{\rm grid}$-axis by $\phi_{\rm bar}$,\footnote{$\phi_{\rm bar}$ is
  slightly different from the observed angle of the bar relative to
  the major axis $\phi_{\rm bar,obs}$, i.e.,\
  $\phi_{\rm bar}\geq\phi_{\rm bar,obs}$. Indeed, approximating a bar
  by a straight line, there is a simple geometrical relationship
  between the two angles:
  $\tan\phi_{\rm bar,obs}=\tan\phi_{\rm bar}\cos i$, although in
  practice the difference between the two angles is reduced by the
  3D structure of the bar. We use $\phi_{\rm bar}$
  instead of $\phi_{\rm bar,obs}$ here for simplicity.} and inclined
with respect to the $X_{\rm grid}$-axis (i.e.,\ the major axis of the
galaxy) by $i$, where $\phi_{\rm bar}$ ranges from $0\degr$ to
  $180\degr$ in steps of $5\degr$ and $i$ ranges from $0\degr$ to $90\degr$
  also in steps of $5\degr$. We generally assume that the disk major
axis is aligned with the $X$-axis of the grid
(PA$_{\rm disk}=90\degr$), although we discuss the case where it is
not in \S~\ref{subsec:recast}.

After rotation, the simulated disk is placed at a distance of
$100$~Mpc, typical of the MaNGA survey sample galaxies
\citep[][]{bun2015}. Assuming a stellar mass-to-light ratio
$M/L=2$~${\rm M}_\sun/{\rm L}_{\sun,r}$ at $r$ band, we then convert
the projected mass to an $r$-band flux (or equivalently an $r$-band
surface brightness $\Sigma$) map, with an original fine sampling
(i.e.,\ spaxel size) of $\approx0\farcs08$ or $40$~pc.

These finely binned flux maps are then convolved by a Gaussian PSF of
variable width to mimic different seeings (i.e.,\ different observing
conditions). Our adopted PSF full widths at half maximum (FWHMs) range
from $0\arcsec$ to $5\arcsec$ in steps of $0\farcs5$, corresponding 
respectively to extremely good (e.g.\ adaptive optics) and bad
seeings. After convolution, to generate realistic mock images, we
further bin all the flux maps to match the Sloan Digitized Sky Survey
(SDSS) imaging survey sampling size, and add the SDSS mean sky
background and noise, resulting in a final coarse sampling (i.e.,\
spaxel size) of $0\farcs4$ or $200$~pc at $D=100$ Mpc
\citep{gunn2006}.

The creation of the required velocity fields is analogous to that of
the flux maps, except for the convolution step. Indeed, unlike mass
(i.e.,\ flux), mean velocity is not a conserved quantity
under convolution, but momentum is. Using the original fine grid, we
therefore first calculate the mean LOS velocity $V_{||}$ in each
spaxel and then multiply the resulting finely binned velocity map by
the associated finely binned flux map, resulting in a finely binned
map of the quantity $V_{||}\Sigma$. It is this map that we then
convolve by the PSF, rather than the finely binned velocity map. The
convolved finely binned $V_{||}\Sigma$ map is then more coarsely
(re)binned as before and divided by the associated coarsely binned
(and convolved) $\Sigma$ map to obtain the desired properly convolved
coarsely binned $V_{||}$ map (see Eqs.~\ref{eq:seeing-sb} and
\ref{eq:seeing-v}).

\begin{figure}[!t]
  \centering
  \includegraphics[width=85mm]{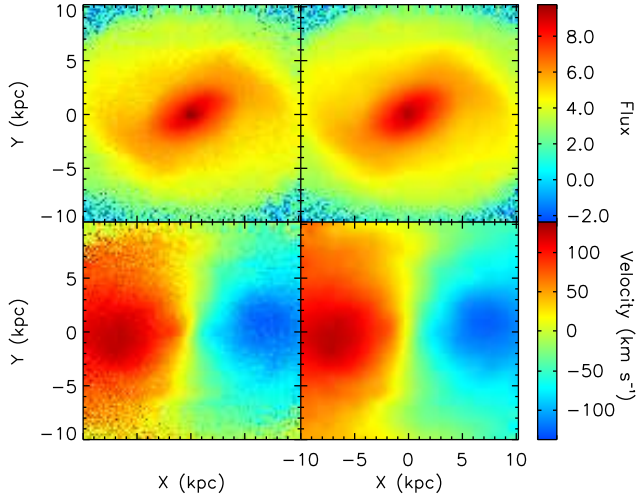}
  \caption{Mock surface brightness (top) and mean
    velocity (bottom) maps, for a relative bar angle
    $\phi_{\rm bar}=45\degr$, an inclination $i=45\degr$,
    no binning, and two different seeings of $0$ (left)
    and $2\arcsec$ (right).}
  \label{fig:mock_img}
\end{figure}

In the end, we thus have mock flux (or surface brightness) and
mean velocity maps with a range of relative bar angles
$\phi_{\rm bar}$, inclinations $i$, and spatial resolutions (PSFs),
all of which have a spaxel size of $200$~pc ($0\farcs4$
  at a distance of $100$~Mpc). Figure~\ref{fig:mock_img} shows
examples of mock surface brightness and mean velocity
maps with $\phi_{\rm bar}=45\degr$, $i=45\degr$, and two different
seeings.

\subsection{Construction of Pseudo-slits}
\label{subsec:pseudo-slits}

Pseudo-slits are positioned according to the projected disk
orientation and bar length in the mock images. Most pseudo-slits are
located within the bar region (in terms of their offset $Y$ from the
galaxy major axis), but as the influence of a bar extends beyond its
extent, we also locate a few extra pseudo-slits beyond the bar, to
test whether these slits still yield accurate measurements of
$\Omega_{\rm p}$. Moreover, the slit widths are varied from $0\farcs4$ to
$2\arcsec$ in steps of $0\farcs4$, to investigate the slit width
influence on the measurements. To obtain an independent
  measurement for each slit, adjacent pseudo-slits touch, but are not 
  overlapped with, each other.

\begin{figure}[!t]
  \centering
  \includegraphics[width=85mm]{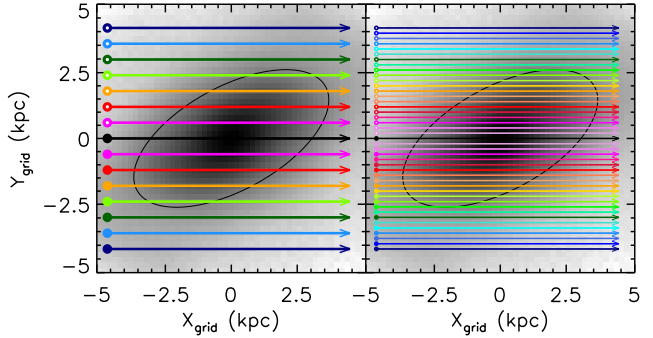}
  \caption{Examples of pseudo-slits schematically overlaid on mock
    images with $\phi_{\rm bar}=45\degr$, $i=45\degr$, and
      no binning. The width of the pseudo-slits is $1\farcs2$ in the
    left panel and $0\farcs4$ in the right panel. The colored straight
    lines schematically represent pseudo-slits with different offsets
    from the major axis (i.e.,\ different $Y_{\rm grid}$
    positions). Arrows indicate the positive direction of the
    pseudo-slits, while the open and filled circles denote
    pseudo-slits with respectively positive and negative offsets from
    the major axis, respectively.}
  \label{fig:pseudo-slit}
\end{figure}

Figure~\ref{fig:pseudo-slit} schematically shows how pseudo-slits are
overlaid on mock images, for both the configuration used in most of
the tests shown ($\phi_{\rm bar}=45\degr$, $i=45\degr$, and a slit
width of $1\farcs2$) and a configuration with narrower $0\farcs4$
slits.

\begin{figure}[!t]
  \centering
  \includegraphics[width=85mm]{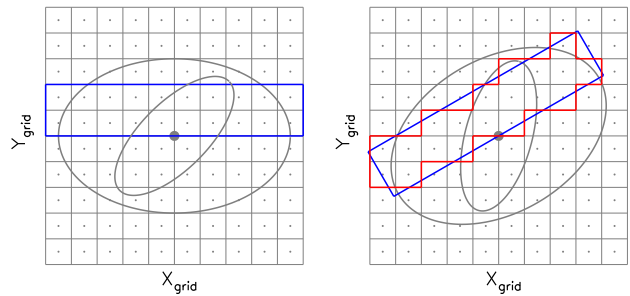}
  \caption{Schematic representations of pseudo-slits with different
    angles between the disk major axis and the data/grid axes
    (PA$_{\rm disk}$). Gray grids and dots show respectively the
    spaxels and their centers, respectively. The large and small gray ellipses show
    respectively the (projected) disk and bar, respectively. Blue rectangles show
    the ideal, perfectly rectangular and symmetric pseudo-slits, which
    should always be parallel to the disk major axis for TW
    measurements. These are easily constructed in the left panel,
    where the disk major axis is aligned with the data/grid axes, but
    are awkward to construct in the right panel, where the disk is at
    an intermediate angle. The red rectangle in the right panel shows
    an irregularly shaped and asymmetric pseudo-slit, comprising only
    those spaxels whose center falls within the desired (perfectly
    rectangular) blue slit.}
  \label{fig:toy-slits}
\end{figure}

In our study, as by construction the galaxy major axis is always
aligned with the $X_{\rm grid}$-axis, we generally use
easily constructed, perfectly rectangular horizontal slits, which are
symmetric with respect to the minor axis (see the left panel of
Fig.~\ref{fig:toy-slits}). However, as in practice the galaxy major
axis (and thus the pseudo-slits) may be misaligned with respect to the
axes of the data/grid, we also create pseudo-slits with irregular
(zigzagged) asymmetric shapes, comprising only those spaxels whose
center falls within the desired (perfectly rectangular) slit (see the
right panel of Fig.~\ref{fig:toy-slits}). We then use those
irregularly shaped pseudo-slits to test their effects (and the
potential effects of recasting the grid) on TW $\Omega_{\rm p}$
measurements in \S\ref{subsec:recast}.

\subsection{Determining $X_{\rm c}$ and $V_{\rm sys}$}
\label{subsec:XcVsys}

Equation~\ref{eq:tw-method}(a) suggests that an exact determination of the
minor axis position ($X_{\rm c}$) and the galaxy systemic velocity
($V_{\rm sys}$) is very important for individual slit measurements and
the TW averaging method. The center and mean velocity  of
our simulated galaxy are initially well determined (and by
construction set to $0$), but thereafter the galaxy can shift slightly
during its evolution. It is thus preferable to determine the value of
$X_{\rm c}$ and $V_{\rm sys}$ from the chosen snapshot of the
simulation itself. In fact, by measuring the mean position and
velocity of our simulated galaxy in increasingly large volumes
(centered on $0$), we see that they are indeed nearly $0$ up to a
radius of $\approx10$~kpc, but that significant offsets (like those
discussed below in Fig.~\ref{fig:XcVsys}) exist when considering
particles farther out, presumably because of ``rogue''particles and
asymmetries in the galaxy outer parts. Since most of the slits
considered in our analyses do reach those distances, it appears
essential to measure $X_{\rm c}$ and $V_{\rm sys}$ empirically from
the simulation data.

Of course, an analogous determination of $X_{\rm c}$ and $V_{\rm sys}$
is likely to be essential for observational data, as they are unlikely
to be known a priori and will need to be determined from the
observational data themselves (i.e.,\ a cube in the case of IFS data)
or from ancillary imaging and spectroscopic data \citep[see,
e.g.,][]{bur1999}.

\begin{figure}[!t]
  \centering
  \includegraphics[width=85mm]{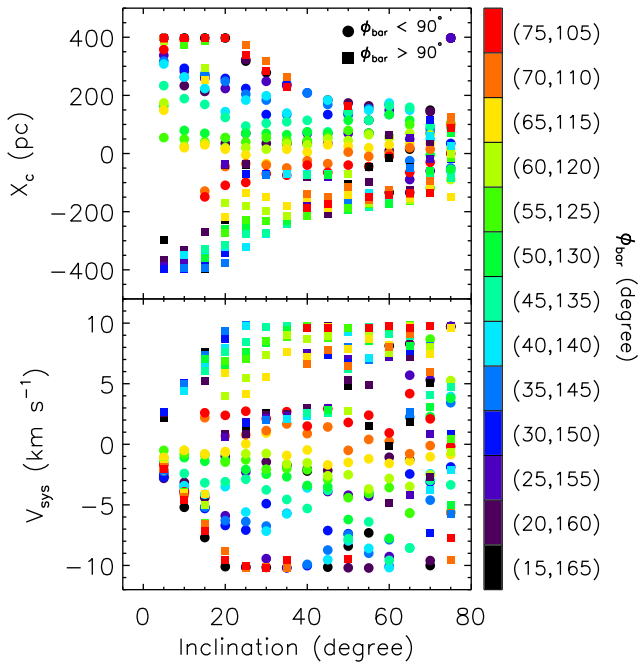}
  \caption{Minor axis position offset $X_{\rm c}$ (top panel) and
    systemic velocity offset $V_{\rm sys}$ (bottom panel) as a
    function of the bar orientation $\phi_{\rm bar}$ and inclination
    $i$ for a seeing of $2\arcsec$, no binning, and
      $\delta$PA$_{\rm disk}=0\degr$, as determined by minimizing the
    sum of the squares of the differences (of $\Omega_{\rm p}$
    measurements) of opposite slit pairs. Results for
      $\phi_{\rm bar}<90\degr$ and $\phi_{\rm bar}>90\degr$ are shown
      as filled circles and squares, respectively.}
  \label{fig:XcVsys}
\end{figure}

Equation~\ref{eq:tw-method}(a) suggests that if $X_{\rm c}$ and/or $V_{\rm sys}$
are wrong, slits with the same offset but on opposite sides of the
galaxy major axis will yield $\Omega_{\rm p}$ measurements that are
systematically (and increasingly) biased in opposite ways (i.e.,\
respectively larger and smaller than the truth). To measure
$X_{\rm c}$ and $V_{\rm sys}$ (for each bar orientation
$\phi_{\rm bar}$, disk inclination $i$, and spatial resolution), we
thus first measure $\Omega_{\rm p}$ for multiple slits (inner slits
only; see \S~\ref{subsec:out_slits}) and for a wide range of possible
$X_{\rm c}$ and $V_{\rm sys}$ values and then select the $X_{\rm c}$
and $V_{\rm sys}$ (and associated $\Omega_{\rm p}$) that minimize the
sum of the squares of the differences of opposite slit pairs (thus
ensuring that the two slits in each pair yield $\Omega_{\rm p}$
measurements that are as similar to each other as possible).

We first let $X_{\rm c}$ and $V_{\rm sys}$ vary over a very wide range
of values for every $(\phi_{\rm bar},i)$ pair. Most $X_{\rm c}$ and
$V_{\rm sys}$ thus selected are small, as expected, but for extreme
values of $\phi_{\rm bar}$ and $i$ the $X_{\rm c}$ and $V_{\rm sys}$
selected are nearly always at the extreme of the ranges allowed (no
matter how large these are), and they often rapidly switch in an
opposite manner from positive to negative values (suggesting that
$X_{\rm c}$ and $V_{\rm sys}$ are somewhat degenerate, as expected
from Eq.~\ref{eq:tw-method}(a)). Since we will later discard those extreme
$(\phi_{\rm bar},i)$ pairs as inappropriate for $\Omega_{\rm p}$
measurements (see \S~\ref{subsec:iphi}), in practice we restrict
$X_{\rm c}$ and $V_{\rm sys}$ to smaller ranges appropriate for most
$(\phi_{\rm bar},i)$ pairs ($-400~pc\le X_{\rm c}\le400$~pc and
$-10$~km~s$^{-1}\le V_{\rm sys}\le10$~km~s$^{-1}$).

Figure~\ref{fig:XcVsys} shows the resulting $X_{\rm c}$ and
$V_{\rm sys}$ offsets (with respect to $0$) as a function of
$\phi_{\rm bar}$ and $i$ for a seeing of $2\arcsec$, no binning, and
no position angle misalignment. Except for extreme values of
$\phi_{\rm bar}$ and $i$, the systemic velocity offsets are nearly
always small (generally $\la8$~km~s$^{-1}$) and show no obvious
trend. The results are similar for the minor axis position offsets,
which can be large for extreme $(\phi_{\rm bar},i)$ pairs but are
otherwise small (generally $\la1$ pixel and always $\la2$ pixels) and
again show no clear trend. The increased minor axis position offsets
at small $i$ are simply due to the deprojection (amplifying small
offsets).

\begin{figure}[!t]
  \centering
  \includegraphics[width=85mm]{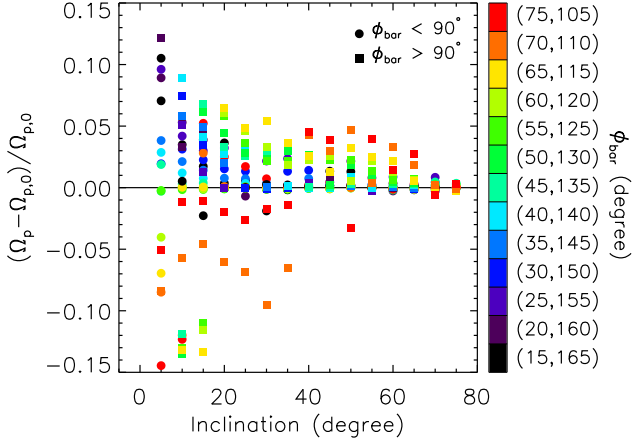}
  \caption{Fractional changes in $\Omega_{\rm p}$ measurements when
    applying $X_{\rm c}$ and $V_{\rm sys}$ offsets (compared to
    measurements with no offset, $\Omega_{\rm p,0}$), as a function of
    the bar orientation $\phi_{\rm bar}$ and inclination $i$ for a
    seeing of $2\arcsec$, no binning, and
      $\delta$PA$_{\rm disk}=0\degr$. Results for
      $\phi_{\rm bar}<90\degr$ and $\phi_{\rm bar}>90\degr$ are shown
      as filled circles and squares, respectively. Except for extreme
    $(\phi_{\rm bar},i)$ pairs, the changes are always $\la3\%$.}
  \label{fig:Omega}
\end{figure}

Importantly, Figure~\ref{fig:Omega} shows that changes in
$\Omega_{\rm p}$ when applying $X_{\rm c}$ and $V_{\rm sys}$ offsets
are small (compared to measurements with no offset), 
  $\la3\%$ excluding extreme $(\phi_{\rm bar},i)$ pairs, as expected
from the fact that the simulation was centered to start with. Of
course, the changes in the case of observational data are likely to be
more significant, as a result of the poorer initial guesses for
$X_{\rm c}$ and $V_{\rm sys}$.

\section{Results and Discussion}
\label{sec:results}

\subsection{Integration along the Slits and Convergence Test}
\label{subsec:convergence}

Ideally, the integrals in Equation~\ref{eq:tw-method}(a) should range from
$-\infty$ to $+\infty$, but this range is necessarily limited in real
observations. As the integration limits (i.e.,\ the length of each
pseudo-slit) increase, measurements of $\Omega_{\rm p}$ based on
Equation~\ref{eq:tw-method}(a) should  asymptotically converge and
reach the true $\Omega_{\rm p}$ value at the (projected) radius where
the disk becomes axisymmetric. It is thus very important to establish
the minimum pseudo-slit length (i.e.,\ the minimum spatial coverage)
required for accurate TW measurements.

\begin{figure}[!t]
  \centering
  \includegraphics[width=80mm]{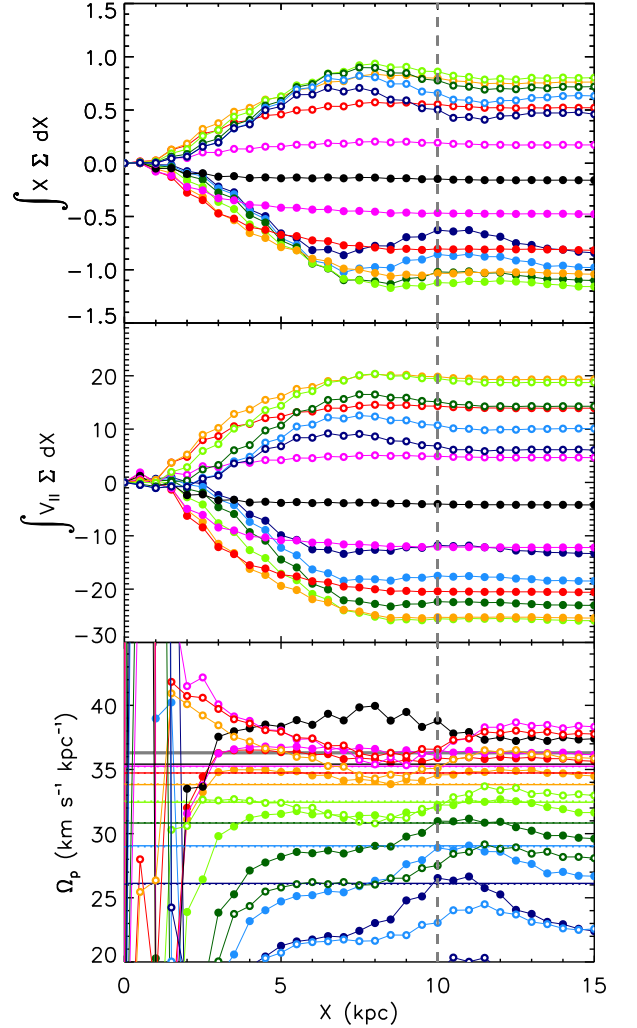}
  \caption{Example convergence test. From top to bottom, the panels
    show distance (i.e.,\ projected radius) profiles of the quantities
    $\int V_{||}\,\Sigma\,dX$, $\int X\,\Sigma\,dX$, and
    $\Omega_{\rm p}$ (see Eq.~\ref{eq:tw-method}(a)), constructed by gradually
    increasing the slit (half-)length (i.e.,\ the integration
    limits). For display purposes, the profiles in the upper and
    middle panels also are normalized by the total flux in each
    slit. Data points with different colors show pseudo-slits with
    different offsets from the simulated galaxy major axis (increasing
    from magenta to dark blue; see Fig.~\ref{fig:pseudo-slit}), with
    open and filled circles showing slits with the same offset but on,
    respectively, the positive and negative side of the major axis (see
    Fig.~\ref{fig:pseudo-slit}). The thick gray solid horizontal line
    shows the intrinsic bar pattern speed of the simulated galaxy
    ($\Omega_{\rm p,int}=36.3$~km~s$^{-1}$~kpc$^{-1}$). The other
    dotted and solid colored horizontal lines indicate the associated
    weighted bar pattern speeds from Equation~\ref{eq:naive}, for,
    respectively, positive and negative offsets from the major
    axis. The gray dashed vertical lines indicate a distance of
    $10$~kpc, where by convention we take our $\Omega_{\rm p}$
    measurements. The mock data set used in this test has
    $\phi_{\rm bar}=45\degr$, $i=45\degr$, $2\arcsec$ seeing,
    $1\farcs2$ slit width, no binning, and
    $\delta$PA$_{\rm disk}=0\degr$.}
  \label{fig:profiles}
\end{figure}

To establish this, we plot in Figure~\ref{fig:profiles} distance
(i.e.,\ projected radius) profiles of the quantities
$\int V_{||}\,\Sigma\,dX$ (the numerator of Eq.~\ref{eq:tw-method}(a)),
$\int X\,\Sigma\,dX$ (the denominator of Eq.~\ref{eq:tw-method}(a)), and
$\Omega_{\rm p}$ (the ratio of the numerator and denominator of
Eq.~\ref{eq:tw-method}(a)  divided by the known $\sin i$),
constructed by gradually increasing the (half-)slit length (i.e.,\ the
integration limits). Multiple pseudo-slits are shown, with different
offsets from the disk major axis, for a mock data set with
$\phi_{\rm bar}=45\degr$, $i=45\degr$, $2\arcsec$ seeing, $1\farcs2$
slit width, no binning, and
  $\delta$PA$_{\rm disk}=0\degr$.

As alluded to above, for ease of interpretation and
comparison to the known intrinsic bar pattern speed
$\Omega_{\rm p,int}=36.3$~km~s$^{-1}$~kpc$^{-1}$, in
Figure~\ref{fig:profiles} and all other figures and measurements in
this section we divide each $\Omega_{\rm p}\sin i$ measurement by
$\sin i$ using the known inclination (i.e.,\ the inclination with which
the simulation was projected). For real observations, the accuracy
with which an $\Omega_{\rm p}\sin i$ measurement can be deprojected
clearly depends on the accuracy of the measured inclination, adding an
additional uncertainty to the $\Omega_{\rm p}$ measurement.

Figure~\ref{fig:profiles} shows that $\int V_{||}\,\Sigma\,dX$ and
$\int X\,\Sigma\,dX$ have converged by a distance of $\approx8$~kpc
($\approx3.3$~$R_{\rm e}$ for our simulated galaxy, much larger than
the bar half-length), where convergence is defined here as being
within $10\%$ of the known true value. However, somewhat surprisingly,
$\Omega_{\rm p}$ converges much earlier, by a distance of
$\approx3$~kpc ($\approx1.3$~$R_{\rm e}$, slightly {\em shorter} than
the projected bar half-length). The convergence distance for
$\int V_{||}\,\Sigma\,dX$ and $\int X\,\Sigma\,dX$ is thus a rather
conservative estimate of the required slit (half-)length, and a much
smaller field coverage is sufficient for reliable TW $\Omega_{\rm p}$
measurements.

While Figure~\ref{fig:profiles} shows the convergence test for a
single $(\phi_{\rm bar},i)$ pair, convergence tests for the entire
range of $\phi_{\rm bar}$ and $i$ yield similar results, i.e.,\ while
the convergence distances of the two integrals taken separately
(numerator and denominator of Eq.~\ref{eq:tw-method}(a)) are always rather large
(and slightly larger for smaller $\phi_{\rm bar}$), the convergence
distance of their ratio (i.e.,\ $\Omega_{\rm p}$) is always much
smaller (and remains essentially unchanged).
\citet{guo2019} also tested the influence of the slit (half-)length on TW 
measurements and found that $\Omega_{\rm p}$ distance profiles converge by 
$\approx1.2$ times the bar half-length. The convergence distances are similar 
for different inclinations and bar orientations.

Having said that, the convergence distance is likely to differ from
galaxy to galaxy, and $\Omega_{\rm p}$ distance profiles may never
converge in the presence of large-scale asymmetries such as spiral
arms and lopsidedness. We therefore recommend to checking whether the
$\Omega_{\rm p}$ distance profiles converge on a slit-by-slit and
galaxy-by-galaxy basis. Only those slits where $\Omega_{\rm p}$ has
converged yield reliable bar pattern speed measurements.


Here, for simplicity, we adopt as the pattern speed measured for one
slit the value at a distance of $10$~kpc (i.e.,\ the absolute value of
the integration limits and thus the half-slit length), well beyond the
convergence distance of any slit.

\subsection{Pseudo-slit Selection}
\label{subsec:out_slits}

The different colors in Figure~\ref{fig:profiles} show pseudo-slits
with different offsets from the simulated galaxy major axis (or,
equivalently, different $Y_{\rm grid}$ positions), while open and
filled circles show slits with the same offset but on opposite sides of
the major axis. Except for the three outermost pseudo-slits, all other
pseudo-slits (which we will refer to as inner slits) are located
within the bar region. The three panels of Figure~\ref{fig:profiles}
show that the distance profiles of the outer slits have significant
fluctuations, much greater than those of the inner slits. In
particular, the (absolute) values of the profiles can decrease with
radius. Most importantly, the bottom panel of
Figure~\ref{fig:profiles} shows that, as the offset from the major
axis increases, all slits yield a systematically and increasingly
biased measurement of $\Omega_{\rm p}$ (toward lower values), with the
three outer slits significantly worse. This is likely due to the two
transient spiral arms with a lower pattern speed in the outer parts of
the simulated disk (see Fig.~\ref{fig:radial_ps}). Indeed, beyond the
bar ends, the contribution of the bar to the pattern becomes less
significant and the spiral arms begin to dominate. It is thus
important to quantify this effect to ascertain the reliability of TW
$\Omega_{\rm p}$ measurements using increasingly offset pseudo-slits.

To investigate the effects of the spiral arms on the bar pattern speed
measurements, we compute a simple weighted measure of $\Omega_{\rm p}$
($\Omega_{\rm p,weighted}$), by weighting the bar and spiral arm
pattern speeds by their likely contributions to the integrals in
Equation~\ref{eq:tw-method}(a). We thus calculate the luminosity within and
beyond the (projected) bar region along each pseudo-slit and then
weight the two pattern speeds accordingly, i.e.,\
\begin{eqnarray}
\Omega_{\rm p,weighted}\equiv\frac{\Sigma_{\rm bar}\,\Omega_{\rm p,bar}+\Sigma_{\rm spiral}\,\Omega_{\rm p,spiral}}{\Sigma_{\rm bar}+\Sigma_{\rm spiral}}\,\,\,,
\label{eq:naive}
\end{eqnarray}
where
$\Omega_{\rm p,bar}=\Omega_{\rm p,int}=36.3$~km~s$^{-1}$~kpc$^{-1}$,
$\Omega_{\rm p, spiral}=20.7$~km~s$^{-1}$~kpc$^{-1}$ (see
Fig.~\ref{fig:radial_ps}), and $\Sigma_{\rm bar}$ and
$\Sigma_{\rm spiral}$ are the fluxes respectively within and beyond
the bar region, respectively. The bar region is defined here as the elliptical
region with an ellipticity equal to the maximum bar ellipticity
($\epsilon_{\rm bar,max}=0.47$) and a semi-major-axis radius equal not
to the bar half-length ($a_{\rm bar}=4.6$~kpc), but rather to the
radial extent of the main feature with a pattern speed equal to that
of the bar in Figure~\ref{fig:radial_ps} ($8$~kpc).


The resulting weighted pattern speeds are shown as horizontal lines in
the bottom panel of Figure~\ref{fig:profiles}, appropriately colored
for each pseudo-slit (dotted and solid lines denote pseudo-slits with,
respectively, positive and negative offsets from the major axis; see
Fig.~\ref{fig:pseudo-slit}). As expected, the figure shows that
$\Omega_{\rm p,weighted}$ systematically decreases with increasing
major axis offset.  Most importantly, $\Omega_{\rm bar,weighted}$
roughly agrees with the TW measurement for all the slits, and it
slowly tends to
$\Omega_{\rm p,spiral}\approx21$~km~s$^{-1}$~kpc$^{-1}$ for the three
outer slits. This indicates that pseudo-slits located outside the bar
region will not yield accurate TW $\Omega_{\rm p}$ measurements (as
outer spiral arms increasingly bias the measurements), in turn
suggesting that only pseudo-slits located within the bar region should
be used.

\begin{figure*}[!t]
  \centering
  \includegraphics[width=150mm]{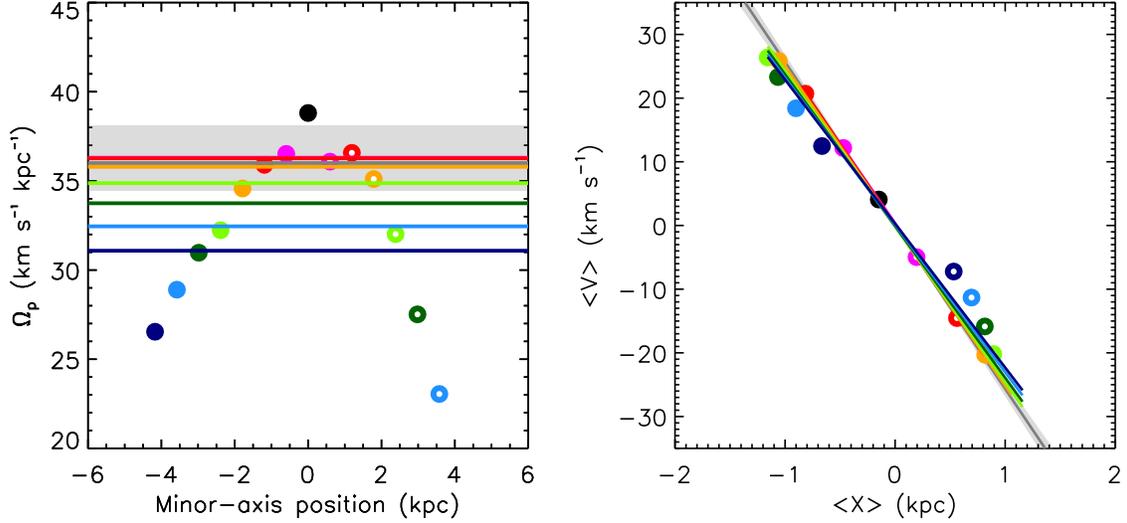}
  \caption{Individual and average bar pattern speed measurements using
    the TW averaging (left) and the TW fitting (right) method. The
    data points are as in Figure~\ref{fig:profiles}. The gray solid
    lines show the intrinsic pattern speed of the bar, and all
    measurements within the pale gray regions have relative errors
    $\le5\%$. Each colored line in the left (right) panel
    shows the average pattern speed (the fit to the $<V>$
    vs. $<X>$ data points) of all the slits within a given offset
    from the galaxy major axis (i.e.,\ all slits within a given
    minor axis position), identified by the large circles of the same
    color. The mock data set used in this test has
    $\phi_{\rm bar}=45\degr$, $i=45\degr$, $2\arcsec$ seeing,
    $1\farcs2$ slit width, no binning, and
      $\delta$PA$_{\rm disk}=0\degr$.}
  \label{fig:comp-slits}
\end{figure*}

Figure~\ref{fig:comp-slits} shows a comparison of the TW
$\Omega_{\rm p}$ measurements for the different slits and for
appropriate slit averages. In the left panel, the TW averaging method
is used, whereby each measurement shown is the average of that for
multiple pseudo-slits. The gray solid line shows the known intrinsic
pattern speed of the bar in the simulated galaxy
($\Omega_{\rm p,int}=36.3$~km~s$^{-1}$~kpc$^{-1}$), while each colored
solid line shows the average pattern speed of all the slits within a
given offset from the galaxy major axis (i.e.,\ all slits within a
given minor axis position $|Y_{\rm grid}|$), identified by the large
circles of the same color. Again, we see that the average
$\Omega_{\rm p}$ decreases as slits with increasingly large offsets
from the major axis are used. In the test shown, the adopted TW
$\Omega_{\rm p}$ measurement would be $>5\%$ lower than the real value
when all pseudo-slits are used, and $\approx5\%$ too low when only the
inner pseudo-slits are used.

The right panel of Figure~\ref{fig:comp-slits} shows an analogous
comparison of the TW $\Omega_{\rm p}$ measurements when the TW fitting
method is used instead. Here $<V>$ and $<X>$ are calculated for each
pseudo-slit, and each adopted $\Omega_{\rm p}$ is obtained from the
fitted slope to the ($<V>$ vs. $<X>$) data points within a given
offset from the galaxy major axis (again color-coded). The results are
consistent with those of the TW averaging method, the fitted slope
decreasing by a similar fraction as slits increasingly offset from the
major axis are incorporated into the fit. The right panel of
Figure~\ref{fig:comp-slits} also shows that the (absolute) values of
$<V>$ and $<X>$ decrease for the outer slits (after peaking for the
pseudo-slits closest to the bar ends).

Compared to the inner slits, we thus see from
Figures~\ref{fig:profiles} and \ref{fig:comp-slits} that the distance
profiles, measurement errors, and $<V>$ -- $<X>$ diagrams of the outer
slits behave differently than those of the inner slits. In addition to
an actual bar (half-)length measurement, these behaviors can help us
to distinguish the inner and outer slits, and thus to select the
appropriate pseudo-slits for a TW $\Omega_{\rm p}$
measurement. Whether using the TW averaging or the TW fitting method,
we therefore recommend using only pseudo-slits located within the bar
region.

\subsection{Spatial Binning}
\label{subsec:spt_bin}
\begin{figure*}[!t]
  \centering
  \includegraphics[width=0.9\textwidth]{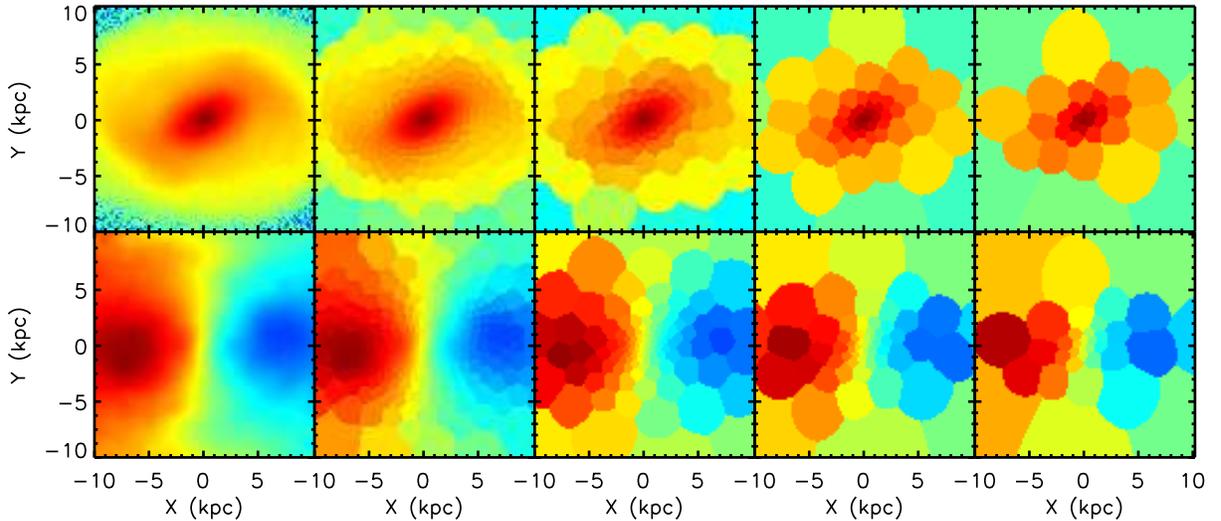}
  \caption{ Examples of binned flux (top) and mean
      velocity (bottom) maps generated using the Voronoi binning
      algorithm. From left to right, the target S/N (number of bins)
      of the Voronoi bins is $0$ ($10101$; no binning), $50$ ($516$),
      $100$ ($142$), $150$ ($64$), and $200$ ($36$). The mock images
      have $\phi_{\rm bar}=45\degr$, $i=45\degr$, and $2\arcsec$
      seeing.}
  \label{fig:bin_maps}
\end{figure*}

To test the effect of spatial binning, we bin the mock images over
multiple spaxels using the commonly used Voronoi binning algorithm of
\citet{cap2003}. Differently from real observations, the noise is
defined here as the Poisson noise of the total flux within each
bin. The spaxels within each bin are labeled, and the surface
brightness and velocity of the bin are then replaced by their
(luminosity-weighted) spatial averages. Figure~\ref{fig:bin_maps}
shows binned flux and mean velocity maps with different S/N
thresholds. Given the necessarily different  S/N definitions, it is
not possible to directly compare the S/N estimates of real
observations with ours, so we use instead the total number of bins
within the field of view to characterize the severity of the binning
applied (more severe binning, or equivalently higher S/N thresholds,
leading to fewer and larger bins). We tested the sensitivity of both
the TW averaging method and the TW fitting method to binning, and both
yield consistent results.

\begin{figure*}[!t]
  \centering
  \includegraphics[width=1.0\textwidth]{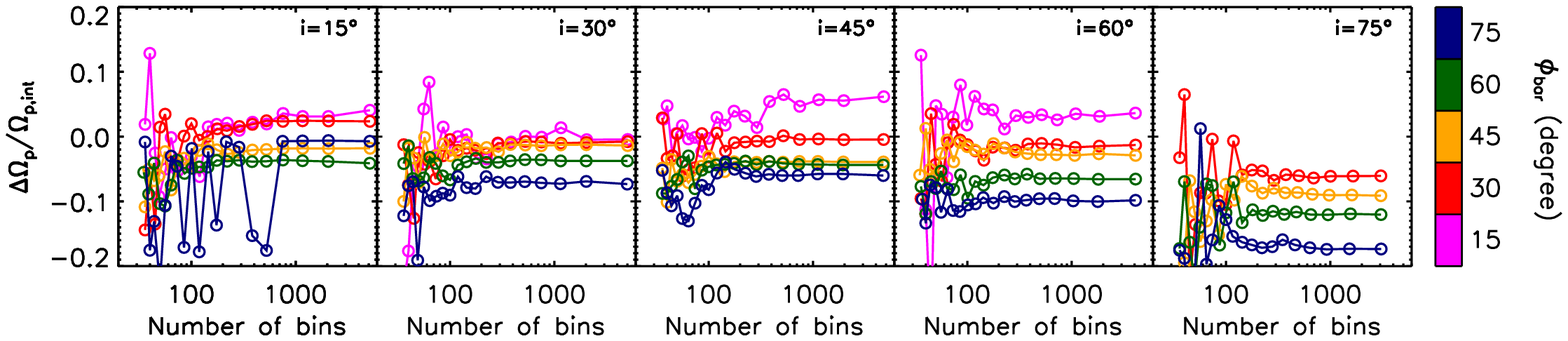}
  \caption{ Spatial binning test. The relative
      $\Omega_{\rm p}$ error is shown as a function of the number of
      bins for different inclinations $i$ (from left to right) and bar
      angles $\phi_{\rm bar}$ (colors). The bar pattern speeds
      $\Omega_{\rm p}$ are measured using the TW averaging method. All
      the mock images used in this test have an FOV of
      $20\times20$~kpc$^2$, $2\arcsec$ seeing, $1\farcs2$ slit width,
      and $\delta$PA$_{\rm disk}=0\degr$.}
  \label{fig:bin_test}
\end{figure*}

Figure~\ref{fig:bin_test} shows the relative error of the bar pattern
speed measurement ($\Delta\Omega_{\rm p}/\Omega_{\rm p,int}$) as a
function of the number of bins for the TW averaging method, for a
range of viewing angles. The FOV of the mock images is
$20\times20$~kpc$^2$ in all cases. As expected, when the number of
bins is large (here $\ga200$, i.e.,\ little binning or low S/N
threshold), spatial binning has no effect on the $\Omega_{\rm p}$
measurements (independently of the viewing angle). Interestingly, the
relative $\Omega_{\rm p}$ error does not increase monotonically with
the number of bins when the number of bins is decreased; it changes
abruptly when the number of bins is sufficiently small (here
$\la200$). This is probably a direct consequence of the importance of
the bins' positions and shapes to the TW integrals. The results of the
TW integrals will be very different for pseudo-slits intersecting,
e.g.,\ a few small bin centers versus the boundaries of several large
bins. Overall, Figure~\ref{fig:bin_test} shows a general trend of
increasing (absolute value of the) relative error
$\Delta\Omega_{\rm p}/\Omega_{\rm p,int}$ with decreasing number of
bins (i.e.,\ more aggressive binning or higher S/N thresholds) when
the number of bins is sufficiently small. The bar pattern speed
$\Omega_{\rm p}$ is then almost always underestimated, by as much as
$15\%$ for the most severe binning considered.

From Figure~\ref{fig:bin_test}, we can draw the conclusion that binned
IFS data with a number of bins smaller than $\approx200$ in an FOV of
$20\times20$~kpc$^2$ should be discarded for TW $\Omega_{\rm p}$
measurements.

\subsection{Grid Recasting}
\label{subsec:recast}

As shown in the right panel of Figure~\ref{fig:toy-slits}, when the
major axis of the disk is misaligned from the data/grid axes, the
integrations along the (pseudo-)slits required for TW measurements
(see Eq.~\ref{eq:tw-method}(a)) are nontrivial. Naively, one can simply
integrate (i.e.,\ sum) along an irregularly shaped slit comprising only
those spaxels whose center falls within the desired (perfectly
rectangular) slit (see the red polygon in the right panel of
Fig.~\ref{fig:toy-slits}). Alternatively, one can recast (i.e.,\
regrid) the grid to align the galaxy major axis with the
$X_{\rm grid}$-axis (see the blue rectangle in the left panel of
Fig.~\ref{fig:toy-slits}), or equivalently carry out a more complex
summation along the slit (see the blue rectangle in the right panel of
Fig.~\ref{fig:toy-slits}).

To quantify the potential biases associated with summing along
irregularly shaped slits and/or the recasting process, we create new
mock data sets where the disk major axis is misaligned from the
$X_{\rm grid}$-axis. Specifically, after projection, we further rotate
the model counterclockwise, so that the angle between the disk major
axis and the data/grid axes (PA$_{\rm disk}$) ranges from $0\degr$ to
$90\degr$ in steps of $5\degr$, this for all $\phi_{\rm bar}$ and $i$.

To simulate the effects of recasting, we subsequently also recast
(i.e.,\ rotate and regrid) all the above mock data sets to align the
disk major axis with the $X_{\rm grid}$-axis. To achieve this, we
apply a bilinear interpolation to the surface brightness ($\Sigma$)
maps directly. Analogously to the convolutions described in
\S~\ref{subsec:mocks}, however, we first multiply the velocity fields
$V_{||}$ by their associated surface brightness maps $\Sigma$ to
obtain $V_{||}\Sigma$ maps. We then apply the bilinear interpolation
to these $V_{||}\Sigma$ maps and divide the recast $V_{||}\Sigma$
maps by their associated recast $\Sigma$ maps, to finally obtain the
desired recast velocity field $V_{||}$. All the recast mock surface
brightness maps and velocity fields then have the disk major axis
aligned with the $X_{\rm grid}$-axis, and the pseudo-slits constructed
are perfectly rectangular, symmetric with respect to the minor axis,
and parallel to the $X_{\rm grid}$-axis (see the left panel of
Fig.~\ref{fig:toy-slits}).

\begin{figure*}[!t]
  \centering
  \includegraphics[width=0.95\textwidth]{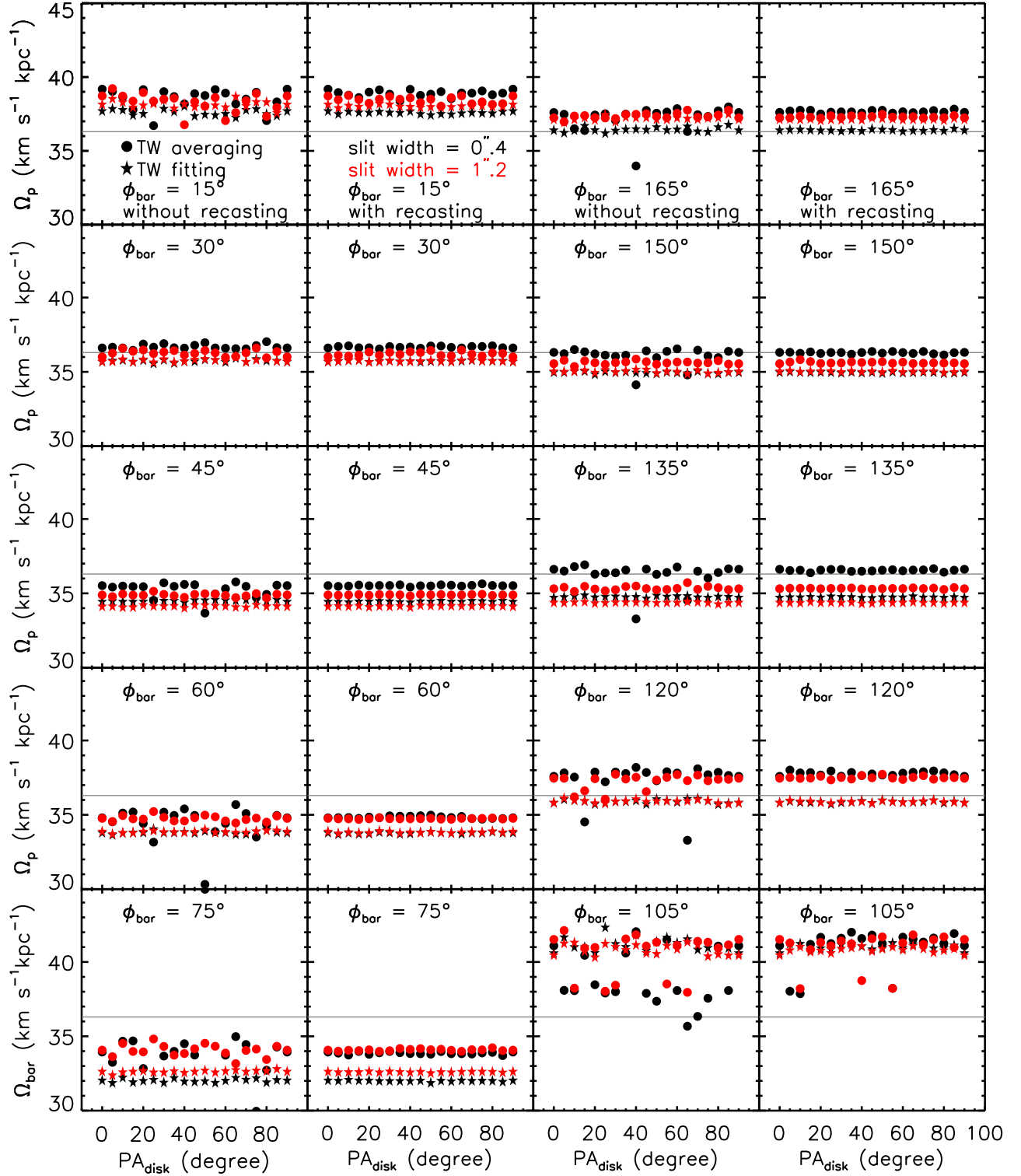}
  \caption{Grid recasting test. TW $\Omega_{\rm p}$ measurements are
    shown as a function of PA$_{\rm disk}$, the angle between the
    galaxy major axis and the data/grid axes (specifically
    $Y_{\rm grid}$), for data sets without (first and third
      columns) and with (second and fourth columns) grid
    recasting. No recasting is required for PA$_{\rm disk}=0$ and/or
    $90\degr$. From top to bottom and left to right,
      $\phi_{\rm bar}$ increases from $15\degr$ to $165\degr$ in steps of
      $15\degr$. Measurements using the TW averaging and fitting
    method are shown, respectively, as filled circles and stars. The
    gray solid horizontal lines show the intrinsic $\Omega_{\rm p}$ of
    the simulated galaxy
    ($\Omega_{\rm p,int}=36.3$~km~s$^{-1}$~kpc$^{-1}$). The mock
    data sets used in this test have $i=45\degr$, $2\arcsec$ seeing,
    $0.4$ (black data points) and $1\farcs2$ (red data points) slit
    widths, no binning, $\delta$PA$_{\rm disk}=0\degr$,
    and a range of $\phi_{\rm bar}$.}
  \label{fig:recast-comp}
\end{figure*}

TW $\Omega_{\rm p}$ measurements are then carried out on the two sets
of simulated data, those recast and those with irregularly shaped
slits. These $\Omega_{\rm p}$ measurements are then compared to each
other and to the known intrinsic bar pattern speed of the simulated
galaxy ($\Omega_{\rm p,int}=36.3$~km~s$^{-1}$~kpc$^{-1}$) in
Figure~\ref{fig:recast-comp}, for mock data sets with $i=45\degr$, a
range of $\phi_{\rm bar}$, $2\arcsec$ seeing, $0.4$ and
  $1\farcs2$ slit widths, no binning, $\delta$PA$_{\rm disk}=0\degr$,
and for both the TW averaging and TW fitting methods. Measurements
carried out without recasting (i.e.,\ with irregularly shaped slits)
and with recasting (i.e.,\ perfectly rectangular slits) are shown
 side by side for comparison.

Comparing the naive integrations along irregularly shaped slits and
recast measurements (i.e.,\  each pair of panels in
Fig.~\ref{fig:recast-comp}), Figure~\ref{fig:recast-comp} shows that
the two sets of measurements are consistent. However, the recast
measurements always show less scatter. Indeed, the relative difference
between the measurements
$\Delta\equiv|\Omega_{\rm p,max}-\Omega_{\rm p,min}|/\Omega_{\rm
  p,int}$ is then $\la2\%$ for both narrow and wide slits, where
$\Omega_{\rm p,max}$ and $\Omega_{\rm p,min}$ are, respectively, the
maximum and minimum derived $\Omega_{\rm p}$ at a given
$\phi_{\rm bar}$. The relative difference $\Delta$ can be as large as
$7\%$ for narrow and wide slits before the recasting
process. Simulated galaxies at different inclinations show the same
trends. This suggests that naive integrations with irregularly shaped
slits do affect TW $\Omega_{\rm p}$ measurements, presumably due to
the asymmetries introduced (and removed by recasting the grid, so that
the only asymmetries left are those due to the bar itself, as required
by TW). Recasting the grid (or, equivalently, carrying out a more
complex summation along the slits) should thus help to reduce
measurement uncertainties.

Interestingly, Figure~\ref{fig:recast-comp} shows that, for any given
$\phi_{\rm bar}$, the TW $\Omega_{\rm p}$ measurements do not
systematically depend on PA$_{\rm disk}$, which simply introduces a
little bit of scatter in the measurements. This is reassuring and
confirms the suggestion above that a misalignment of the galaxy major
axis from the data/grid axes (and potential associated recasting
effects) does not significantly impact TW $\Omega_{\rm p}$
measurements with IFSs. However, grid recasting does help to decrease
the scatter of TW measurements with PA$_{\rm disk}$.

Figure~\ref{fig:recast-comp} also reveals that the TW fitting and TW
averaging methods generally share the same trends. However, the TW
fitting method systematically yields slightly lower TW
$\Omega_{\rm p}$ measurements than the TW averaging method, the
difference increasing with increasing $\phi_{\rm bar}$. This is
somewhat surprising, as one would naively expect the TW fitting method
to be superior to the TW averaging method. We do not fully understand
this tendency, but it seems to be related to the aforementioned fact
that the $\Omega_{\rm p}$ measurements decrease with increasing offset
from the major axis. Indeed, the inner slits near the ends of the bar
(i.e.,\ the green datapoints in the right panel of
Fig.~\ref{fig:comp-slits}) yield both the lowest $\Omega_{\rm p}$
estimates and the greatest (absolute) values of $<V>$ and $<X>$, thus
effecting the greatest leverage (down) on $\Omega_{\rm p}$ when
fitting the straight line. This in turn yields lowered estimates
compared to ones with effectively equal weight for all
datapoints. This effect becomes more acute for larger $\phi_{\rm bar}$
as the intersection of the bar and the slit decreases, and it can be
minimized by neglecting the slits near the ends of the bar.

In any case, most importantly here, even after recasting
Figure~\ref{fig:recast-comp} shows that the scatter across the
measurements is most significant at small $\phi_{\rm bar}$ (i.e.,\
$\phi_{\rm bar}\la15\degr$; both $\phi_{\rm bar}\la15\degr$ and
$75\degr\la\phi_{\rm bar}\la105\degr$ before
recasting). The relative difference $\Delta\approx2\%$ when
$\phi_{\rm bar}\ga15\degr$, but it can be as large as
$7\%$ for $\phi_{\rm bar}\la15\degr$ and $11\%$ for
  $75\degr\la\phi_{\rm bar}\la105\degr$ before recasting. This
suggests that bar pattern speed measurements of galaxies with
disadvantageous $\phi_{\rm bar}$ will have large uncertainties, and
that these galaxies should be excluded from observational samples. In
\S~\ref{subsec:iphi}, we shall therefore explore the optimal range of
both $\phi_{\rm bar}$ and $i$ for sample selection.

\subsection{Disk Position Angle}

\subsubsection{Pattern Speed Uncertainties}
\label{subsec: pa_error}

\begin{figure*}[!t]
  \centering
  \includegraphics[width=1.0\textwidth]{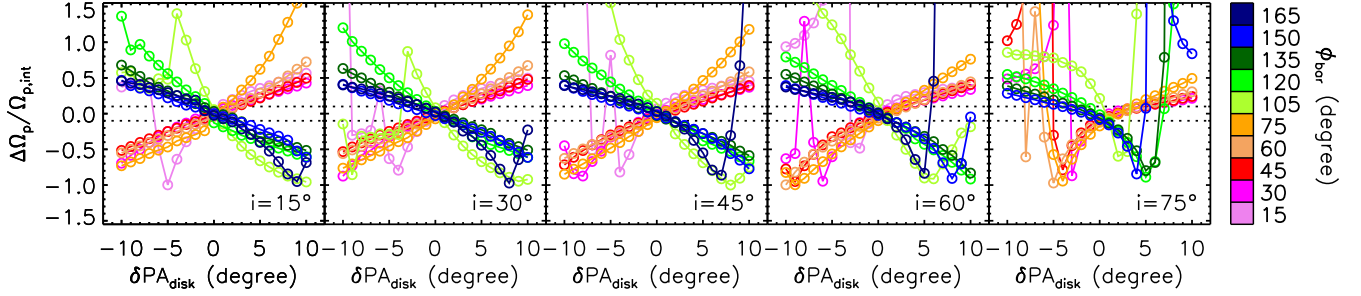}
  \caption{ Position angle test. The relative
      $\Omega_{\rm p}$ error is shown as a function of the position
      angle misalignment angle $\delta$PA$_{\rm disk}$ for different
      inclinations $i$ (from left to right) and bar angles
      $\phi_{\rm bar}$ (colors). The black dotted horizontal lines
      mark relative errors of $\pm10\%$. The bar pattern speeds
      $\Omega_{\rm p}$ are measured using the TW averaging method. The
      mock data sets used in this test have a $2\arcsec$ seeing,
      $1\farcs2$ slit width, and no binning.}
  \label{fig:pa_err_test}
\end{figure*}

To quantify the effects of misalignments ($\delta$PA$_{\rm disk}$) of
the (pseudo-)slits with respect to the true position angle of the disk
(PA$_{\rm disk}$), we must create new mock images. As described in
\S~\ref{subsec:mocks}, the simulated bar is normally first aligned
with the data $X_{\rm grid}$-axis, and the disk is then rotated about
the $Z_{\rm grid}$-axis counterclockwise by $\phi_{\rm bar}$ and
about the $X_{\rm grid}$-axis by $i$. After these two rotations, the
disk major axis lies along the $X_{\rm grid}$-axis by construction.
To introduce a misalignment of the pseudo-slits, we further rotate the
disk about the $Z_{\rm grid}$-axis by an angle $\delta$PA$_{\rm disk}$
(but continue to assume that the disk major axis is along the
$X_{\rm grid}$-axis for the purpose of the TW calculations). For these
tests, $i$ and $\phi_{\rm bar}$ range from $15\degr$ to $165\degr$ in steps
of $15\degr$ and $\delta$PA$_{\rm disk}$ ranges from $-10\degr$ to $10\degr$
in steps of $1\degr$. By convention (and as in D03),
$\delta$PA$_{\rm disk}>0$ when the true disk major axis is rotated
anticlockwise. For $\phi_{\rm bar}\la90\degr$, the additional
rotation increases the angle between the bar and the assumed major
axis (i.e.,\ the $X_{\rm grid}$-axis), while for
$\phi_{\rm bar}\ga90\degr$ the additional rotation decreases that
angle. We carry out this test using both the TW averaging method and
the TW fitting method, but the results are very similar. The relative
$\Omega_{\rm p}$ errors introduced by position angle misalignments are
thus shown in Figure~\ref{fig:pa_err_test} for the TW averaging method
only.

Figure~\ref{fig:pa_err_test} confirms the result of D03 that not only are position
angle misalignments the most important factor potentially
affecting TW $\Omega_{\rm p}$ measurements, but their effect is also
systematic. Indeed, Figure~\ref{fig:pa_err_test} suggests that the
(absolute value of the) relative error
$\Delta\Omega_{\rm p}/\Omega_{\rm p,int}$ is $\approx35\%$ for all
inclinations $i$ and most bar angles $\phi_{\rm bar}$ when the
misalignment angle is as small as $5\degr$. When
$75\la\phi_{\rm bar}\la105\degr$, the relative $\Omega_{\rm p}$ errors
can be very large for even smaller $\delta$PA$_{\rm disk}$. In fact,
when $\delta$PA$_{\rm disk}\approx10\degr$,
$\Delta\Omega_{\rm p}/\Omega_{\rm p,int}$ can reach $100\%$. This test
thus suggests that the measured $\Omega_{\rm p}$ can be severely
overestimated when the angle between the bar and the assumed disk
major axis is wrongly increased, while it can be severely
underestimated when that angle is decreased.


We note that D03 also examined the effects of position angle
misalignment on TW $\Omega_{\rm p}$ measurements. He used a simulated
barred galaxy with $i=45\degr$ and $\phi_{\rm bar}=30\degr, 45\degr$ and $60\degr$ (his
Figure~9), which can be directly compared with our tests. The
results are analogous, except that our relative errors are slightly
smaller, probably due to differences in the simulated galaxies.

\subsubsection{Empirical Diagnostic}
\label{sec: pa_diagnostic}

Given the importance of identifying the true disk position angle
PA$_{\rm disk}$ to avoid the substantial systematic biases just
described, it would be very useful to have an empirical diagnostic of
it, i.e.,\ a method to identify or measure the true disk position angle
from the data themselves. In this subsection, we therefore explore
such empirical diagnostics, unfortunately only with partial success.

We test three different parameters (scatters) to quantify the
``goodness'' of the position angle used: (a) the sum of the squares of
the differences of opposite slit pair measurements, (b) the sum of the
squares of the differences of all slit measurements with respect to
the mean (i.e.,\ the scatter across all pseudo-slit measurements), and
(c) the sum of the previous two parameters. We define the goodness of a
given position angle as
$\left[1-\frac{(\sigma-\sigma_{\rm min})}{(\sigma_{\rm
      max}-\sigma_{\rm min})}\right]$, where $\sigma$ is the given
measure of scatter evaluated at that position angle and
$\sigma_{\rm min}$ and $\sigma_{\rm max}$ are, respectively, the minimum
and maximum scatter measured within the entire position angle ranged
probed. This goodness ranges from $0$ to $1$ by definition, with the
maximum value ($1$) defining what naively would appear to be the most
appropriate disk position angle to use. We compare the goodness of a
range of $\delta$PA$_{\rm disk}$ for a number of $(\phi_{\rm bar},i)$
pairs (and for each of the three scatters defined) in Figure~\ref{fig:
  pa_diagnostic}.

\begin{figure*}[!t]
  \centering
  \includegraphics[width=1.0\textwidth]{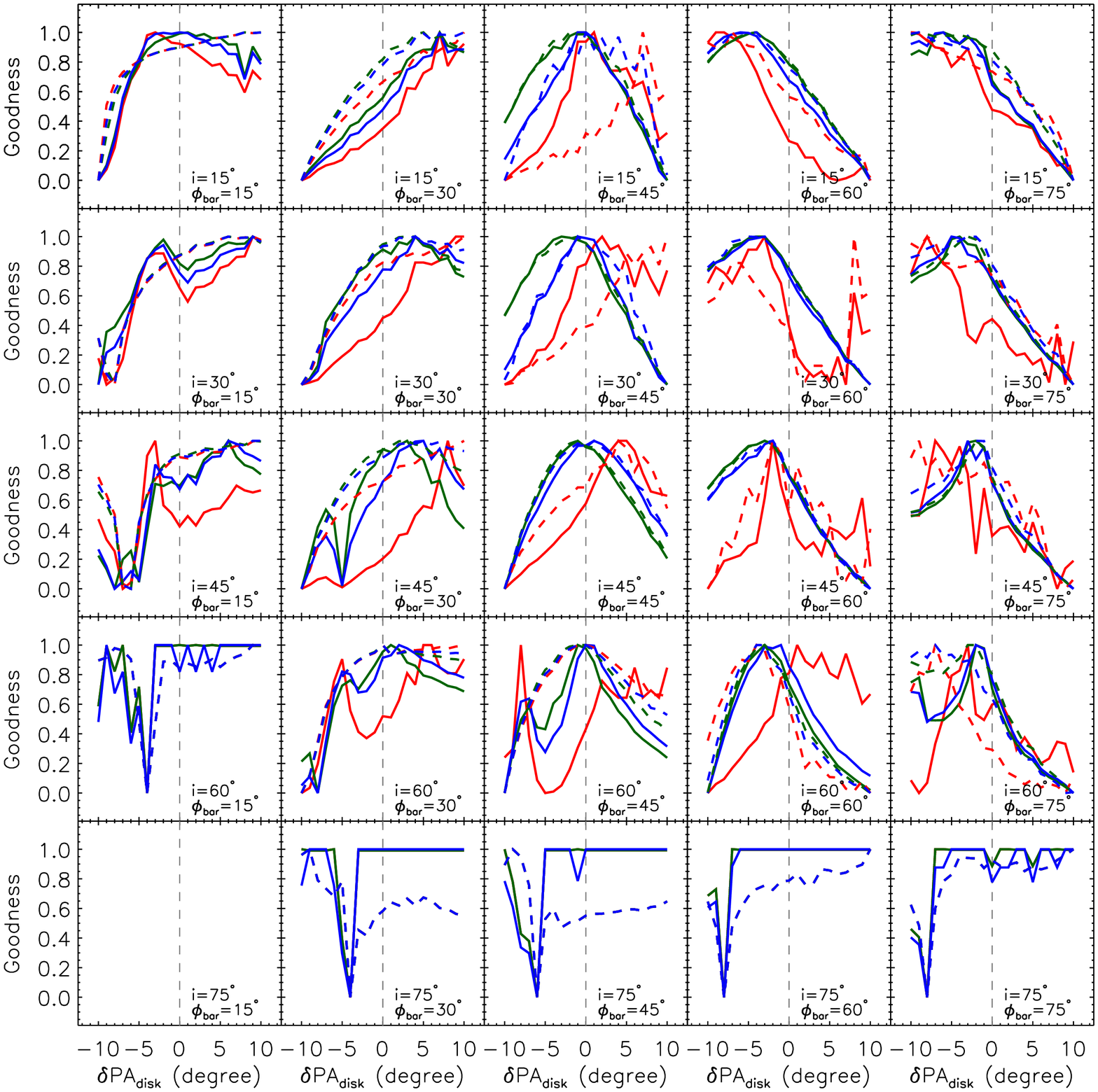}
  \caption{Empirical diagnostics of the true disk position
      angle. The goodness of the parameter is shown as a function of
      $\delta$PA$_{\rm disk}$ for a range of bar angles
      $\phi_{\rm bar}$ (increasing from left to right in steps of
      $15\degr$) and inclinations $i$ (increasing from top to bottom
      in steps of $15\degr$), this for the three parameters discussed
      in the text: sum of the squares of the differences of opposite
      slit pair measurements (a, red), sum of the squares of the
      differences of all slit measurements with respect to the mean
      (b, green), and sum of the previous two parameters (c,
      blue). The goodness ranges from $0$ to $1$ by definition. Solid
      (dashed) lines show the goodness when applying (
      not applying) $X_{\rm c}$ and $V_{\rm sys}$ offsets.}
  \label{fig: pa_diagnostic}
\end{figure*}

We naively expected the goodness to be maximum ($1$) for
$\delta$PA$_{\rm disk}=0\degr$, thus allowing us to empirically
identify the true position angle when it is not accurately known a
priori. However, Figure~\ref{fig: pa_diagnostic} reveals the situation
to be more complex. First, the maximum of the first parameter tested
(a, red curves) is often not very well defined. Second, even for the
second (b, green curves) and third (c, blue curves) parameters tested,
although a clear maximum is often present {\em near}
$\delta$PA$_{\rm disk}=0\degr$, it is often away from {\em exactly}
$\delta$PA$_{\rm disk}=0\degr$ (nevertheless suggesting that the second
parameter, the scatter across all inner pseudo-slits, is the most
promising diagnostic tested). Indeed, the peaks in the goodness
profiles of the second and third parameters often deviate from
$\delta$PA$_{\rm disk}=0\degr$ by a few degrees ($\la5\degr$) when
$i\la60\degr$ (the goodness profiles for $i=75\degr$ are
uninformative). When $\phi_{\rm bar}=45\degr$, the true disk position
angle is accurately recovered (i.e.,\ the goodness is maximum for
$\delta$PA$_{\rm disk}\approx0\degr$). However, when the bar is close
to the disk major axis ($\phi_{\rm bar}<45\degr$), the suggested disk
position angle is always slightly larger than the truth (i.e.,\ the
goodness is maximum for $\delta$PA$_{\rm disk}>0\degr$). Conversely,
when the bar is close to the disk minor axis
($\phi_{\rm bar}>45\degr$), the suggested disk position angle is
always smaller than the truth (i.e.,\ the goodness is maximum for
$\delta$PA$_{\rm disk}<0\degr$). Given the sensitivity of
$\Omega_{\rm p}$ to a disk position angle error of even a few degrees
(see the previous \S~\ref{subsec: pa_error}), these diagnostics are
simply not good enough.

While showing some promise, the potential empirical diagnostics of the
true disk position angle presented in Figure~\ref{fig: pa_diagnostic}
thus fall short of what is required to ensure bar pattern measurements
devoid of significant PA$_{\rm disk}$-related systematic errors. We
also do not fully understand the systematic dependence of the goodness
adopted on the bar angle $\phi_{\rm bar}$. Further investigations are
thus required to identify a more reliable empirical diagnostic of the
true disk position angle.
  
\subsection{Bar Orientation and Inclination}
\label{subsec:iphi}

As discussed in \S~\ref{subsec:tw-formalism} (and according to
Eq.~\ref{eq:tw-method}(a)), the TW integrals will sum to $\approx0$ if a bar is
nearly parallel ($\phi_{\rm bar}=0\degr$) or perpendicular
($\phi_{\rm bar}=90\degr$) to the major axis of the disk. In addition,
the LOS velocities will be very small (and undistinguishable within
the uncertainties) if a galaxy is close to face-on ($i=0\degr$), while
a bar will be difficult to identify if a galaxy is close to edge-on
($i=90\degr$). It is thus necessary to exclude those galaxies with
disadvantageous $\phi_{\rm bar}$ and $i$ for reliable TW
$\Omega_{\rm p}$ measurements.

There is no specific prescription for the optimal $\Omega_{\rm p}$
ranges, so we use our simulated galaxy with different projections to
identify those optimal ranges here. To do this, we will consider two
quantities. First, the relative difference of each average
$\Omega_{\rm p}$ measurement with respect to the known intrinsic bar
pattern speed ($\Omega_{\rm p,int}=36.3$~km~s$^{-1}$~kpc$^{-1}$),
where the TW average is taken over all individual inner slit
measurements of a given $(\phi_{\rm bar},i)$ pair. Second, the
relative uncertainty of each average $\Omega_{\rm p}$ measurement with
respect to the known intrinsic bar pattern speed, where the
uncertainty is defined as the root mean square (rms) scatter across
all individual inner slit measurements of a given $(\phi_{\rm bar},i)$
pair.

We note that there are very few inner pseudo-slits when
$\phi_{\rm bar}\la10\degr$ (or $\phi_{\rm bar}\ga170\degr$) and/or the
disk is nearly edge-on, as the bar is then nearly parallel to the
major axis of the disk and/or the projected bar length is too
short. By the time $i\approx90\degr$ is reached, measurements are
essentially impossible for any $\phi_{\rm bar}$. In particular, the
uncertainty measure defined above requires at least two slits.

\begin{figure*}[!t]
  \centering
  \includegraphics[width=0.8\textwidth]{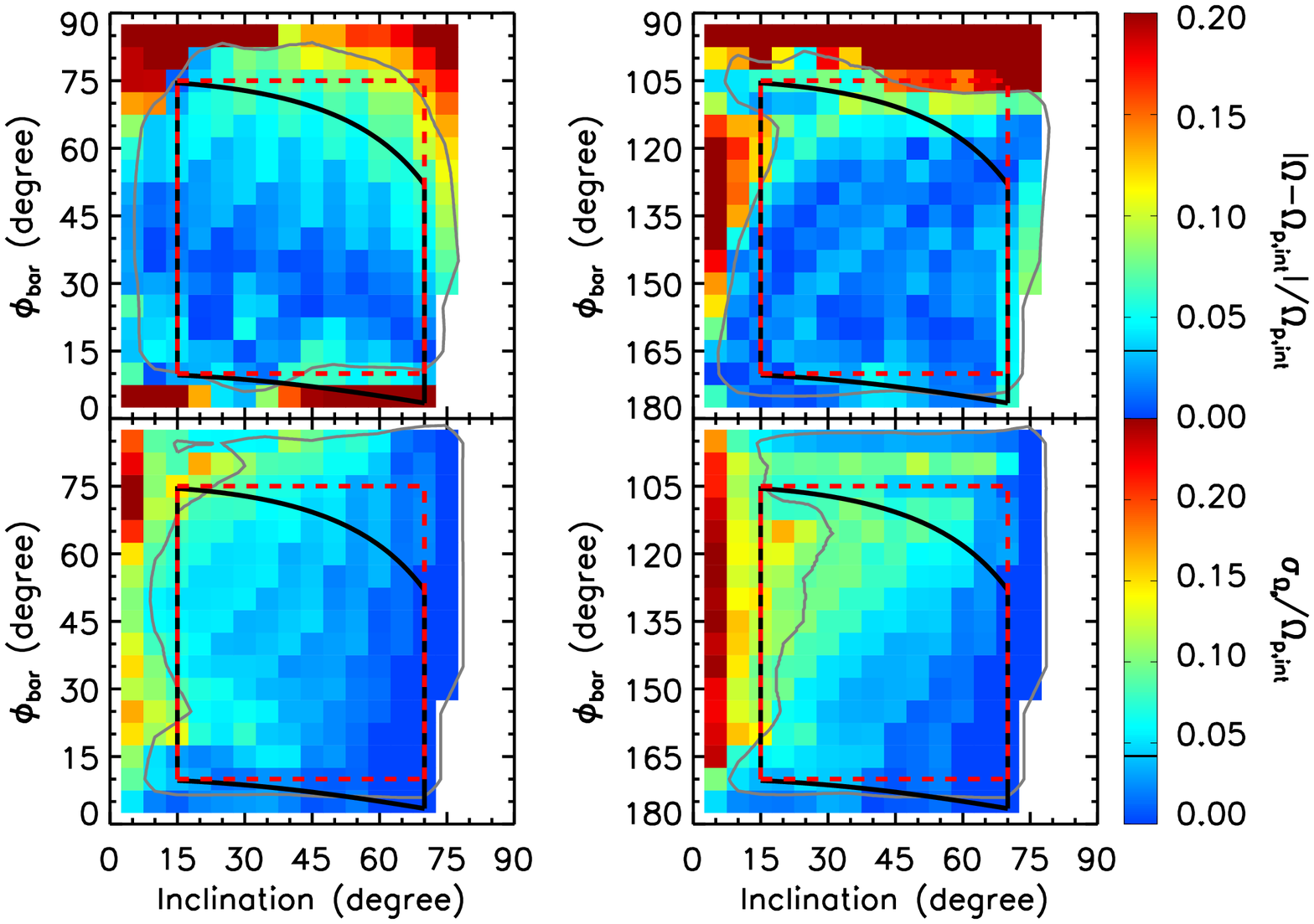}
  \caption{Optimal $\phi_{\rm bar}$ and $i$ range test for
      $\phi_{\rm bar}\le90\degr$ (left panels) and
      $\phi_{\rm bar}\ge90\degr$ (right panels). Colors in the top
    panels show the (absolute value of the) relative
    difference of each average $\Omega_{\rm p}$ measurement (with
    respect to the known intrinsic bar pattern speed
    $\Omega_{\rm p,int}=36.3$~km~s$^{-1}$~kpc$^{-1}$) as a function of
    $\phi_{\rm bar}$ and $i$, where the average is taken over all
    individual inner slit measurements. Colors in the bottom panels
    show the relative uncertainty of each average $\Omega_{\rm p}$
    measurement, defined as the ratio of the rms scatter across all
    individual inner slit measurements to the known intrinsic bar
    pattern speed $\Omega_{\rm
      p,int}=36.3$~km~s$^{-1}$~kpc$^{-1}$. White regions indicate
    $(\phi_{\rm bar},i)$ pairs inappropriate for TW $\Omega_{\rm p}$
    measurements ($i=0\degr$, $\phi_{\rm bar}=0$ or
      $180\degr$, single inner slit). Solid gray lines are $10\%$
    contours. The red dashed box identifies the optimal intrinsic ($\phi_{\rm bar}$, $i$) region (as used throughout this paper), while the black box identifies the optimal observed ($\phi_{\rm bar, obs}$, $i$) region (see Footnote 2). The optimal observed $\phi_{\rm bar,obs}$ range (black box) should be used when selecting samples for TW measurements. The mock
    data sets used in this test have $2\arcsec$ seeing, $1\farcs2$ slit
    width, no binning, $\delta$PA$_{\rm disk}=0\degr$, and
    a range of $\phi_{\rm bar}$ and $i$.}
  \label{fig:i-pa}
\end{figure*}

Figure~\ref{fig:i-pa} shows the (absolute values of the)
$\Omega_{\rm p}$ relative differences and uncertainties derived from
mock data sets with a $2\arcsec$ seeing, $1\farcs2$ slit width,
no binning, $\delta$PA$_{\rm disk}=0\degr$, and a range of
$\phi_{\rm bar}$ and $i$. From the figure, we see that the relative
differences are $\ga10\%$ when $\phi_{\rm bar}\la10\degr$
  and $75\degr\la\phi_{\rm bar}\la105\degr$, or $i\la15\degr$ and
  $i\ga70\degr$. Similarly, the relative uncertainties are $\ga10\%$
  for $75\degr\la\phi_{\rm bar}\la105\degr$ or $i\la15\degr$. In any case,
it is difficult to accurately measure the inclination of a nearly
face-on galaxy, as the disk appears nearly (but may not be
intrinsically perfectly) round, and the deprojection correction to the
velocities (or equivalently the $\Omega_{\rm p}\sin i$ TW measurement)
is very large. The projection of the bar on the minor axis is also too
short to construct multiple pseudo-slits when $\phi_{\rm bar}$ is
small.

\begin{figure}[!t]
  \centering
  \includegraphics[width=85mm]{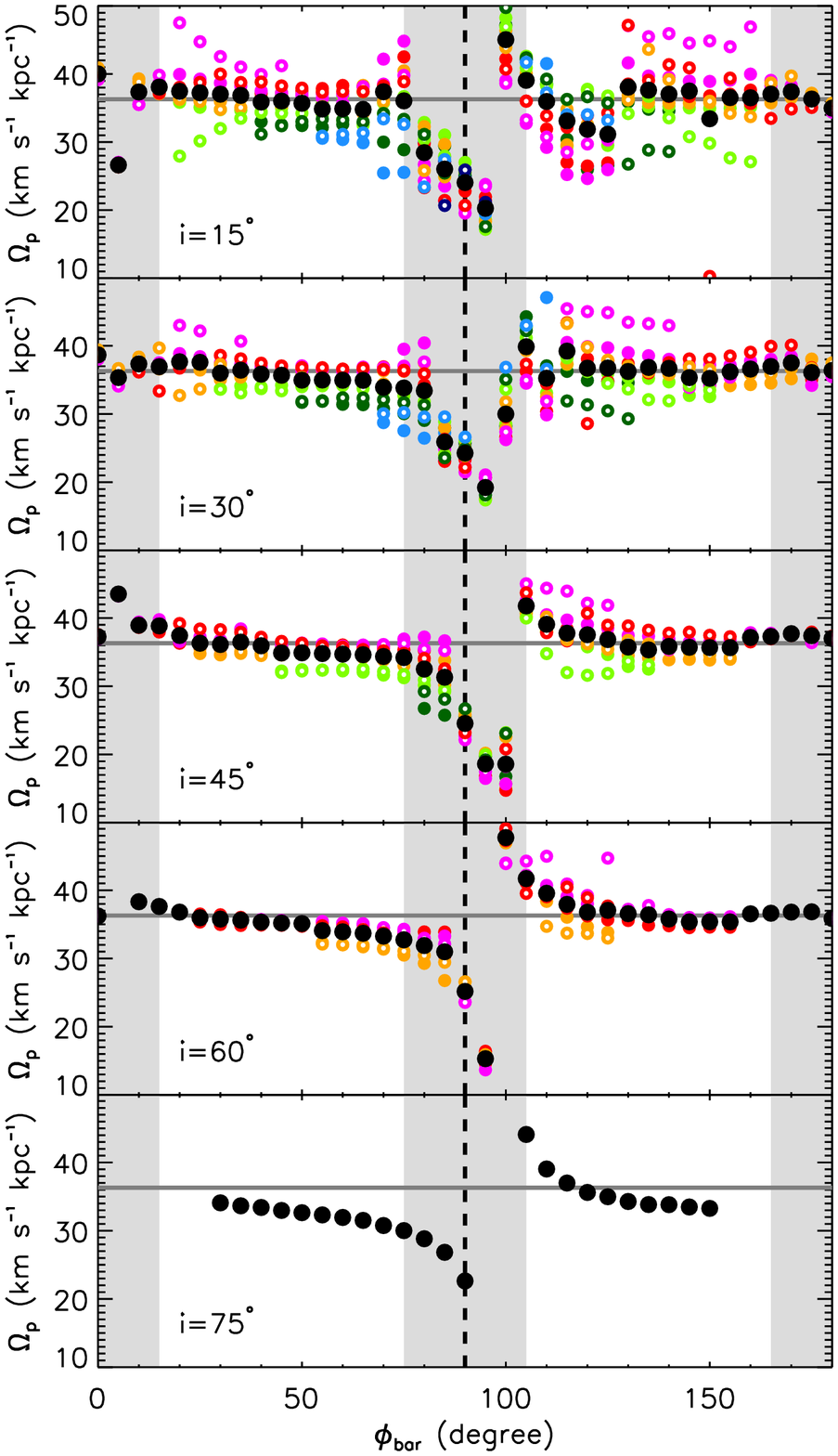}
  \caption{TW $\Omega_{\rm p}$ measurements for individual
      pseudo-slits (and their averages) as a function of
      $\phi_{\rm bar}$ and $i$. As in Figure~\ref{fig:profiles}, data
      points with different colors show pseudo-slits with different
      offsets from the simulated galaxy major axis (increasing from
      magenta to pale blue; see Fig.~\ref{fig:pseudo-slit}), with open
      and filled circles showing slits with the same offset but on the positive and negative side, respectively, of the major axis
      (see Fig.~\ref{fig:pseudo-slit}). The large black filled circles
      show the averages of all the slits. The thick gray solid
      horizontal lines show the intrinsic bar pattern speed of the
      simulated galaxy
      ($\Omega_{\rm p,int}=36.3$~km~s$^{-1}$~kpc$^{-1}$), while the
      shaded gray areas show the bar angles when the bar is near the
      major or the minor axis of the disk. The bar pattern speeds
      $\Omega_{\rm p}$ are measured using the TW averaging method. The
      mock data sets used in this test have a $2\arcsec$ seeing,
      $1\farcs2$ slit width, no binning, and
      $\delta$PA$_{\rm disk}=0\degr$.}
  \label{fig:omega_phibar}
\end{figure}

 As already mentioned, according to Eq.~\ref{eq:tw-method}(a), the TW
  integrals will sum to $\approx0$ if a bar is nearly parallel
  ($\phi_{\rm bar}=0\degr$) or perpendicular
  ($\phi_{\rm bar}=90\degr$) to the major axis of the disk. The
  $\Omega_{\rm p}$ inferred is then expected to have large
  uncertainties. Figure~\ref{fig:i-pa} indeed suggests that the
  relative differences are very large when
  $\phi_{\rm bar}\approx90\degr$, but surprisingly the relative
  differences remain small for $\phi_{\rm bar}\approx0\degr$ (or,
  equivalently, $\phi_{\rm bar}\approx180\degr$). To investigate this,
  we show in Figure~\ref{fig:omega_phibar} the TW $\Omega_{\rm p}$
  measurements for individual pseudo-slits (and their averages), for a
  range of bar angles $\phi_{\rm bar}$ and inclinations
  $i$. Figure~\ref{fig:omega_phibar} reveals a clear systematic
  dependence of the measurements on the bar orientation. The
  relatively good quality of the measurements when
  $\phi_{\rm bar}\approx0\degr$ is confirmed. However, unexpectedly,
  not only are the measurements poor for
  $\phi_{\rm bar}\approx90\degr$, but the curves appear antisymmetric
  about $\phi_{\rm bar}\approx90\degr$ as well. We do not fully
  understand this, but we surmise that it is due to the decreasing
  contribution of the bar to the TW integrals when the angle between
  the bar and the major axis increases (while the influence of the
  spiral arms remains). In any case, TW $\Omega_{\rm p}$ measurements
  should be treated with particular caution when the bar is close to
  the disk minor axis. It might also be that erroneous measurements at
  $\phi_{\rm bar}\approx90\degr$ could explain so-called ultrafast
  bars \citep{guo2019}.


Our tests therefore provide useful guidelines for the selection of
observational samples for TW $\Omega_{\rm p}$ measurements. Indeed,
using Figure~\ref{fig:i-pa}, one can exclude inappropriate
$\phi_{\rm bar}$ and $i$ that would result in significant measurement
biases and/or uncertainties. For example, according to our simulation,
and to keep both of these quantities to $\la10\%$, one should select
barred disk galaxies with $10\degr\la\phi_{\rm bar}\la75\degr$
and $105\degr\la\phi_{\rm bar}\la170\degr$ and
$15\degr\la i\la70\degr$. Measurements are particularly
  inaccurate when the bar is close to the disk minor axis (see also
  Figure~\ref{fig:omega_phibar}). It is worth noting that the optimal $\phi_{\rm bar}$ range is quoted in the intrinsic face-on values. In Figure~\ref{fig:i-pa}, we also show the optimal observed $\phi_{\rm bar, obs}$ range with a black box when projection effects are taken into account (see Footnote 2). The optimal observed $\phi_{\rm bar, obs}$ range (black box) should be used when selecting samples for TW measurements.

Having said that, our tests are based on a single simulated barred
galaxy, and the relative differences and uncertainties shown in
Figure~\ref{fig:i-pa} would likely change slightly for other
simulations/galaxies. In any case, as noted before, even favorable
$(\phi_{\rm bar},i)$ pair TW $\Omega_{\rm p}$ measurements can still
suffer from large biases because of other large-scale asymmetric
structures such as spiral arms and lopsidedness. Our recommendation
thus remains to always look at the $\Omega_{\rm p}$ distance profile
for each slit and apply the convergence test discussed in
\S~\ref{subsec:convergence} and Figure~\ref{fig:profiles}.

\subsection{Slit Width}
\label{subsec:slit}

\begin{figure}[!t]
  \centering
  \includegraphics[width=85mm]{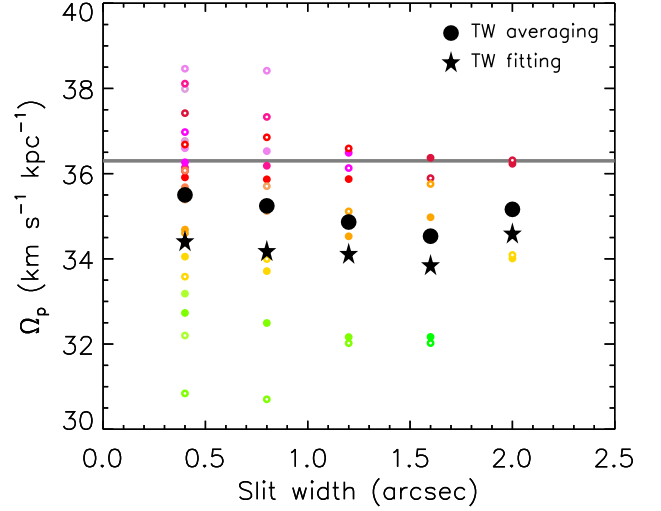}
  \caption{Slit width test. TW $\Omega_{\rm p}$ measurements are shown
    as a function of the width of the pseudo-slit used. Large black
    filled circles and stars show average measurements (inner slits
    only) using, respectively, the TW averaging and the TW fitting
    method. Small colored circles show $\Omega_{\rm p}$ measurements
    from individual pseudo-slits with different offsets from the major
    axis (see Fig.~\ref{fig:pseudo-slit}). The gray solid horizontal
    line shows the intrinsic $\Omega_{\rm p}$ of the simulated galaxy
    ($\Omega_{\rm p,int}=36.3$~km~s$^{-1}$~kpc$^{-1}$). The mock
    data sets used in this test have $\phi_{\rm bar}=45\degr$,
    $i=45\degr$, $2\arcsec$ seeing, no binning,
      $\delta$PA$_{\rm disk}=0\degr$, and variable slit widths.}
  \label{fig:slit-width}
\end{figure}

Our mock data sets span pseudo-slit widths of $0\farcs4$ to $2\arcsec$ in
steps of $0\farcs4$, with no overlap or gap between adjacent
pseudo-slits. The number of slits that can be used thus decreases with
increasing slit width (as the absolute size of the simulated galaxy is
fixed). The bar pattern speeds measured from those individual
pseudo-slits are shown in Figure~\ref{fig:slit-width}, along with the
averages using the TW averaging and TW fitting methods applied to the
inner slits only.

Overall, all the average measurements shown in
Figure~\ref{fig:slit-width} are in agreement with each other to within
$\approx3\%$, well within any reasonable uncertainties
(defined by, e.g., the scatter across individual measurements for each
slit width). The slit width therefore has no significant effect on the
measured average $\Omega_{\rm p}$.

As was already pointed out from Figure~\ref{fig:recast-comp} (see also
Fig.~\ref{fig:comp-slits}), we again note that the TW fitting method
yields slightly and systematically lower measurements than the TW
averaging method, but again this is only by a few percent and within
reasonable uncertainties.

We also note that the scatter between individual measurements (i.e.,\
the $\Omega_{\rm p}$ measured from individual slits) increases
drastically for very narrow slits (i.e.,\ slits significantly narrower
than the seeing). As the measured $\Omega_{\rm p}$ varies
systematically with the offset from the galaxy major axis (see
\S~\ref{subsec:out_slits}), this is probably simply due to the wider
slits encompassing (and thus averaging) wider ranges of
$\Omega_{\rm p}$. In any case, overly narrow pseudo-slits should be
avoided for single slit measurements. Importantly, however, the
average measurements remain reliable, so that again we conclude that
the pseudo-slit width has no significant impact on the measurements.

\subsection{Spatial Resolution}
\label{subsec:seeing}

Angular resolution is usually considered very important for
observations, as it determines the ability to see intricate spatial
details in the objects studied. In optical/infrared observations, this
angular resolution is usually quantified by the FWHM of the Gaussian
PSF (i.e.,\ the seeing). In this study, we use spatial and angular
resolution interchangeably, as the distance to our simulated galaxy is
fixed, and we use our simulation to test the influence of spatial
resolution on TW measurements.

\begin{figure}[!t]
  \centering
  \includegraphics[width=85mm]{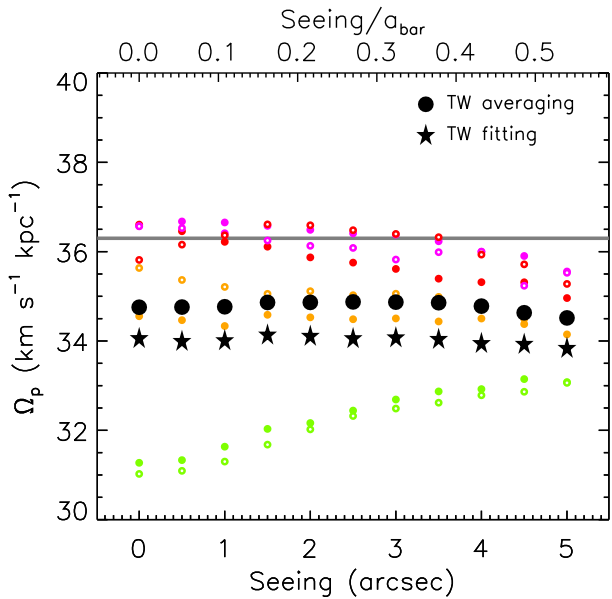}
  \caption{Spatial resolution test. TW $\Omega_{\rm p}$ measurements
    are shown as a function of the spatial resolution (seeing FWHM)
    used. Symbols and lines are as in Figure~\ref{fig:slit-width}. The
    mock data sets used in this test have $\phi_{\rm bar}=45\degr$,
    $i=45\degr$, a $1\farcs2$ slit width, no binning,
      $\delta$PA$_{\rm disk}=0\degr$, and variable seeing.}
  \label{fig:seeing}
\end{figure}

Figure~\ref{fig:seeing} shows TW $\Omega_{\rm p}$ measurements using
mock data sets with seeings ranging from $0\arcsec$ to $5\arcsec$ in steps of
$0\farcs5$, and using both the TW averaging and the TW fitting
method. The average measurements using the TW fitting method are again
slightly lower than those using the TW averaging method. Of more
interest here, however, there is a weak trend of decreasing
$\Omega_{\rm p}$ with increasing seeing for the innermost slits (i.e.,\
the magenta, red, and orange datapoints in Fig.~\ref{fig:seeing}). The
same trend is seen with other mock data sets with different
$\phi_{\rm bar}$ and $i$, and it is easily understood from the trend
of decreasing $\Omega_{\rm p}$ with increasing offset from the galaxy
major axis described in \S~\ref{subsec:out_slits}. Indeed, with
increasingly bad seeing, particles at increasingly large
$Y_{\rm grid}$ (or, equivalently, increasingly large major axis
offsets) influence the observed surface brightnesses and velocities
within any given pseudo-slit (see Eqs.~\ref{eq:seeing-sb} and
\ref{eq:seeing-v}). The measured $\Omega_{\rm p}$ thus decreases
slightly but systematically with seeing. The opposite is true for the
last inner slits near the ends of the bar (i.e.,\ the green datapoints
in Fig.~\ref{fig:seeing}), presumably for the same reason, i.e.,\ with
increasingly bad seeing particles at increasingly small $Y_{\rm grid}$
contribute to the measurements, and in consequence $\Omega_{\rm p}$
systematically increases with increasing seeing.

Most important here, however, is that the average measurements are essentially
independent of seeing. Indeed, even for an unrealistically large
seeing of $5\arcsec$, the relative error
($\Omega_{\rm p,5^{\prime\prime}}-\Omega_{\rm
  p,0^{\prime\prime}}$)/$\Omega_{\rm p,int}$ of the average
measurements is $<1\%$, where $\Omega_{\rm p,5^{\prime\prime}}$ and
$\Omega_{\rm p,0^{\prime\prime}}$ are the measured bar pattern speeds
for respectively $5\arcsec$ and $0\arcsec$ seeing, respectively, and
$\Omega_{\rm p,int}=36.3$~km~s$^{-1}$~kpc$^{-1}$ is the known
intrinsic bar pattern speed of the simulation. For a typical MaNGA
seeing of $\approx2\arcsec$, this relative error is also $<1\%$.
Considering realistic uncertainties (again, e.g., the scatter across
individual measurements for each seeing), Figure~\ref{fig:seeing}
suggests that spatial resolution has no significant effect on TW
measurements. The TW method is thus reliable even at low spatial
resolutions, as originally suggested by \cite{tre1984}. Interestingly,
it is thus well suited for poor-seeing backup observing programs.

\subsection{TW Method and IFS Data}

Combining the advantages of imaging and spectroscopy, IFSs are now
available on most telescopes. These instruments, however, have different
characteristics (wavelength coverage, FOV, spectral resolution,
angular sampling, etc.), that make them most appropriate for different
observational strategies and thus scientific goals. More IFSs will
also come online in the near future.

For TW $\Omega_{\rm p}$ measurements, only those IFSs with a
relatively large FOV are appropriate (see
\S~\ref{subsec:convergence}). We therefore select some IFSs with a
large FOV and summarize their characteristics in
Table~\ref{tab:ifs_instruments}. All operate in the optical domain,
with FOV diameters larger than $30\arcsec$ and spectral resolutions
and angular samplings appropriate for dynamical work on nearby
galaxies. Two Fabry-Perot instruments (GHaFaS and FaNTOmM)
specifically target H$\alpha$ emission, thus providing high-quality
spectra for application of the TW method to ionized gas. Their FOVs
are very large, and spectral resolutions very high. A new generation of
imaging Fourier transform spectrographs has similar strengths (SpIOMM
and SITELLE). Instruments mounted on large telescopes and/or
exploiting adaptive optics have particularly high angular samplings
(MUSE, VIMOS, Kyoto3DII). Although a high spatial resolution is not
crucial for TW measurements (see \S~\ref{subsec:seeing}), the data
from these instruments also generally have higher S/N and can be
binned more flexibly, which should lead to higher-quality TW
$\Omega_{\rm p}$ measurements. The SparsePak, DensePak, and PPack
family of fiber-fed IFSs have unusually high spectral resolutions, but
this is unlikely to be a main driver for TW measurements.

\begin{deluxetable*}{lccccr}
  \tablecaption{Characteristics of IFS instruments appropriate for TW
    pattern speed measurements.\label{tab:ifs_instruments}}
  \tablecolumns{6} \tablewidth{0pt} \tablehead{ \colhead{Instrument} &
    \colhead{Wavelength Range} & \colhead{Spectral Res.} &
    \colhead{Spatial Sampling} & \colhead{Field of View} &
    \colhead{Telescope}\\
    \colhead{} & \colhead{(\AA)} & \colhead{} & \colhead{} &
    \colhead{} & \colhead{} } \startdata
  RSS & $4300$ -- $\phantom{,1}8600$ & $300$ -- $9000$ & $0\farcs13$ & $8^{\prime}$ & SALT 10 m \\
  Kyoto3DII & $4000$ -- $\phantom{,1}7000$ & $400$, $7000$ & $0\farcs056$ & $1\farcm9\times1\farcm9$ & Subaru 8.2 m \\
  MUSE & $4650$ -- $\phantom{,1}9300$ & $\approx3000$ & $0\farcs2$ & $1^{\prime}\times1^{\prime}$ & VLT 8 m\\
  VIMOS & $3600$ -- $10,000$ & $200$ -- $2500$ & $0\farcs67$ & $54\arcsec\times54\arcsec$ & VLT 8 m\\
  SAURON & $4500$ -- $\phantom{,1}7000$ & $\approx1500$ & $0\farcs94$ & $41\arcsec\times33\arcsec$ & WHT 4.2 m \\
  GHaFaS & $4500$ -- $\phantom{,1}8500$ & $\approx20,000$ & $0\farcs4$ & $3\farcm4\times3\farcm4$ & WHT 4.2 m \\
  WEAVE$^{*}$ & $3700$ -- $\phantom{,1}9600$ & $5000$, $20,000$ & $1\farcs3$, $2\farcs6$ & $11\arcsec\times12\arcsec$, $78\arcsec\times90\arcsec$ & WHT 4.2 m \\
  SAMI & $3700$ -- $\phantom{,1}9500$ & $1700$ -- $13,000$ & $1\farcs6$ & $15\arcsec$ & AAT 3.9 m \\
  DensePak & $3700$ -- $11,000$ & $5000$ -- $20,000$ & $3\farcs0$ & $30\arcsec\times45\arcsec$ & WIYN 3.8 m \\
  SparsePak & $5000$ -- $\phantom{,1}9000$ & $5000$ -- $20,000$ & $4\farcs7$ & $72\arcsec\times71\farcs3$ & WIYN 3.8 m \\
  SITELLE & $3500$ -- $\phantom{,1}9000$ & $1$ -- $10,000$ & $0\farcs32$ & $11^{\prime}\times11^{\prime}$ & CFHT 3.6 m\\
  PPak & $4000$ -- $\phantom{,1}9000$ & $\approx8000$ & $2\farcs7$ & $74\arcsec\times64\arcsec$ & Calar Alto 3.5 m \\
  VIRUS-P & $3500$ -- $\phantom{,1}6800$ & $\approx850$ & $4\farcs3$ & $1\farcm7\times1\farcm7$ & McDonald 2.7 m \\
  VIRUS-W & $4340$ -- $\phantom{,1}6040$ & $2500$, $6800$ & $3\farcs2$ & $105\arcsec\times75\arcsec$ & McDonald 2.7 m \\
  MaNGA & $3600$ -- $10,400$ & $\approx2000$ & $2\farcs0$ & $12\farcs5$ -- $32\farcs5$ & APO 2.5 m \\
  CHILI$^{*}$ & $3600$ -- $\phantom{,1}7200$ & $900$, $1900$ & $5\farcs6$ & $71\arcsec\times65\arcsec$ & Lijiang 2.4 m\\
  FaNTOmM & $4000$ -- $\phantom{,1}9000$ & $10,000$ -- $60,000$ & $1\farcs6$ & $17^{\prime}$ & OMM 1.6 m\\
  SpIOMM & $3500$ -- $\phantom{,1}9000$ & $1$ -- $25,000$ & $0\farcs5$ & $12^{\prime}\times12^{\prime}$ & OMM 1.6 m \\
  \enddata
  \tablenotetext{*}{Future IFS.}
\end{deluxetable*}

Over the past two decades, the development of IFSs has motivated many
IFS surveys of nearby galaxies, the most important of which are
summarized in Table~\ref{tab:ifs_surveys}. The Spectral Areal Unit for
Research on Optical Nebulae (SAURON) and subsequent Atlas$^{\rm 3D}$
projects using the SAURON IFS pioneered large-scale surveys, observing
a mass- ($K$-band luminosity) and volume-limited sample of $260$
early-type galaxies \citep{cap2011}. The Calar Alto Legacy Integral
Field Area (CALIFA) survey used the PPak IFS to study a sample of
$\approx670$ galaxies selected for their optical size
\citep{san2012}. The VIRUS-P Exploration of Nearby Galaxies (VENGA)
survey observed a sample of $30$ nearby spiral galaxies chosen for
their extensive ancillary multiwavelength data \citep{bla2013}. The
Sydney-Australian-Astronomical-Observatory Multi-object Integral-field
Spectrograph (SAMI) survey is currently targeting $3400$ galaxies
using a multiplexed fiber-fed IFS with two wavelength channels
\citep{croo2012}. The Mapping Nearby Galaxies at APO (MaNGA) project
aims to observe $10,000$ galaxies over a wide range of stellar masses
and environments \citep{bun2015}.


Table~\ref{tab:ifs_surveys} shows that Atlas$^{\rm 3D}$ has the best
spatial sampling but a relatively small spatial coverage and poor
velocity resolution. CALIFA has the largest FOV, helping to ensure
that the optical extent of most galaxies is fully covered (and thus
that the TW integrals converge). The FOV of VENGA is also large, but
its velocity resolution is rather poor for TW measurements. SAMI and
MaNGA have good compromises between FOV and spatial sampling, as well
as between wavelength coverage and velocity resolution. The advantage
of MaNGA is its large sample size, but SAMI covers a wider range of
environments. Both should enable comprehensive studies of the
correlations between bar pattern speeds and other properties of
galaxies. The FOV of MaNGA reaches at least $1.5R_{\rm e}$ for
two-thirds of the sample and at least $2.5R_{\rm e}$ for one-third,
the latter being particularly promising given the convergence test
discussed in \S~\ref{subsec:convergence} (showing that TW
$\Omega_{\rm p}$ measurements typically converge by
$\approx1.3R_{\rm e}$). As a second-generation IFS, the Multi Unit
Spectroscopic Explorer \citep[MUSE;][]{bac2010} provides a unique
combination of large FOV, high spatial sampling, and generally high
S/N, offering new opportunities for accurate $\Omega_{\rm p}$
measurements using the TW method.

\begin{deluxetable*}{lrccccc}
  \tablecaption{Characteristics of relevant IFS
    surveys.\label{tab:ifs_surveys}} \tablecolumns{7} \tablewidth{0pt}
  \tablehead{ \colhead{Survey} & \colhead{Sample Size} &
    \colhead{Field of View} & \colhead{Spatial Elements} &
    \colhead{Wavelength Range} & \colhead{Velocity Res.} &
    \colhead{Angular Res.} \\
    \colhead{} & \colhead{} & \colhead{} & \colhead{} &
    \colhead{(\AA)} & \colhead{($\sigma$; km~s$^{-1}$)} &
    \colhead{(Reconstructed FWHM)} } \startdata
  MaNGA & $10,000$ & $>1.5$, $>2.5$~$R_{\rm e}$ & $19$ -- $127$ & $3600$ -- $10,300$ & $50$ -- $80$ & $\approx2\farcs5$ \\
  WEAVE & $5000$ & $11\arcsec\times12\arcsec$, $78\arcsec\times90\arcsec$ & $37$, $540$ & $3660$ -- $\phantom{,1}9840$ & $15$, $60$ & \nodata\\
  SAMI & $3400$ & $>1$~$R_{\rm e}$ & $61$ & $3700$ -- $\phantom{,1}5700$ & $70$ & $\approx2\farcs2$ \\
  & & & & $6300$ -- $\phantom{,1}7400$ & $30$ & $\approx2\farcs2$ \\
  CALIFA & $667$ & $>2.5$~$R_{\rm e}$ & $331$ & $3700$ -- $\phantom{,1}7000$ & $85$, $150$ & $\approx2\farcs4$ \\
  Atlas$^{\rm 3D}$ & $260$  & $\approx1$~$R_{\rm e}$ & $1431$ & $4800$ -- $\phantom{,1}5380$ & $105$ & \nodata \\
  SIGNALS & $38$ & $11\arcmin\times11\arcmin$ & $\approx4,500,000$ & $3630$ -- $\phantom{,1}6850$ & $60$ & $\approx0\farcs8$ \\
  VENGA & $30$ & $\approx0.7$~$R_{\rm 25}$$^{*}$ & $246$ & $3600$ -- $\phantom{,1}6800$ & $\approx120$ & $\approx5\farcs6$ \\
  \enddata
  \tablenotetext{*}{$R_{\rm
      25}$ is the radius of the
    $25$~mag~arcsec$^{-2}$ isophote in the $R$ band.}
\end{deluxetable*}

\section{Conclusions}
\label{sec:concl}
In this work, we aimed to assess the potential limitations, biases,
and uncertainties of direct bar pattern speed ($\Omega_{\rm p}$)
measurements using the method proposed by \citet{tre1984}, with a
special emphasis on IFS observations. To
achieve this, we used a simple $N$-body simulation of a barred disk
galaxy and created a series of mock data sets with different
alignments of the galaxy major axis with respect to the data/grid axes
(PA$_{\rm disk}$), bar angles with respect to the galaxy major axis
($\phi_{\rm bar}$), disk inclinations ($i$), (pseudo-)slit widths,
spatial resolutions (seeings), spatial binnings, and
position angle misalignments. We summarize our main findings below.

(i) A convergence test is essential to establish the reliability of
any TW $\Omega_{\rm p}$ measurement, whereby a profile of the derived
$\Omega_{\rm p}$ is constructed as a function of the distance along
the slit (i.e.,\ the integration limits of the TW integrals). This
profile should converge to a constant value for any measurement to be
deemed reliable, and it therefore determines the minimum field of view
(or, equivalently, the minimum [half-]slit length) required for
accurate measurements (a projected distance of $1.3$ times the
effective or half-mass radius for our simulation).

(ii) We find that the derived $\Omega_{\rm p}$ is increasingly biased
low as the offset of a (pseudo-)slit from the galaxy major axis
increases, and that only (pseudo-)slits located within the bar region
yield accurate measurements.

Bar pattern speeds measured from slits beyond the bar can be
significantly affected by other large-scale asymmetric structures with
different (generally lower) pattern speeds, such as spiral arms and
lopsidedness. In consequence, when slits near the ends of the bar are
used, the TW fitting method generally yields slightly lower TW
$\Omega_{\rm p}$ measurements than the TW averaging method.

(iii) Bar pattern speed measurements using IFSs are not
  significantly affected by spatial binning unless the bins are
  extremely large and/or few, when TW $\Omega_{\rm p}$ measurements
  can be underestimated by as much as $\approx15\%$. IFS data with a
  number of bins smaller than $\approx200$ in an FOV of
  $20\times20$~kpc$^2$ should not be used for TW measurements.

(iv) We find no significant dependence of the TW
$\Omega_{\rm p}$ measurements on the position angle of
  the disk within the IFS FOV. However, recasting the grid to align
the disk major axis with the data/grid axes does lead to smaller
uncertainties on the measurements (compared to naive integrations
along irregularly shaped slits comprising only those spaxels whose
center falls within the desired perfectly rectangular slits).

(v) We confirm the finding of D03 that the TW method is
  very sensitive to misalignments of the disk position angle (and thus
  the orientation of the pseudo-slits). Position angle misalignments
  lead to TW $\Omega_{\rm p}$ measurement errors that are both the
  largest and are systematic. For a misalignment of as little as
  $5\degr$, the relative $\Omega_{\rm p}$ systematic error can be as
  large as $35\%$. When the angle between the bar and the assumed disk
  major axis is wrongly increased, the inferred $\Omega_{\rm p}$ can
  be severely overestimated, leading, for example, to apparently
  ultrafast bars, conversely when the angle between the bar and
  the assumed disk major axis is wrongly decreased. We unfortunately
  fail to build a sufficiently accurate empirical disk position angle
  diagnostic, but we do provide approximate diagnostics pointing the way
  for future studies.

(vi) Given the limitations introduced by both
observational effects and the TW formalism itself, we determined the
optimal $\phi_{\rm bar}$ and $i$ ranges for accurate TW
$\Omega_{\rm p}$ measurements. For our simulation, the relative biases
and uncertainties can be kept to $\la10\%$ for
$10\degr\la\phi_{\rm bar}\la75\degr$ and
  $105\degr\la\phi_{\rm bar}\la170\degr$ and $15\degr\la i\la70\degr$.
 Measured bar pattern speeds are significantly less
  accurate when the bar is close to the disk minor axis.

(vii) We find that unless the slit width is
significantly smaller than the seeing, the slit width does not affect
TW $\Omega_{\rm p}$ measurements significantly.

(viii) As expected from the TW formalism, the spatial
resolution (i.e.,\ seeing) of the observations does not have a
significant impact on TW $\Omega_{\rm p}$ measurements either.

To improve current TW bar pattern speed measurements, we
  thus suggest to choose targets with appropriate viewing angles
  ($\phi_{\rm bar}$ and $i$), to select the pseudo-slits carefully,
  and to apply the convergence test for every slit in every
  galaxy. Most importantly, only those galaxies with small major axis
  position angle uncertainties should be considered. Minor
  improvements can be obtained by paying attention to the grid
  recasting.

  Our results suggest that it is possible to wrongfully infer
  ultrafast bars if the angle between the bar and the assumed disk
  major axis is overestimated, if the bar is close to the disk minor
  axis, and/or if the FOV is too small for convergence.

Our tests thus provide useful guidelines for future applications of
the TW method to real data, particularly IFS observations. However,
as our tests are based on a single simulation, the exact limitations,
biases, and uncertainties of TW $\Omega_{\rm p}$ measurements will
vary slightly from galaxy to galaxy.

\acknowledgments We thank the referee for the constructive comments
that helped to improve the quality of the paper.
The research presented here is partially supported by
the National Key R\&D Program of China under grant No.\
2018YFA0404501; by the National Natural Science Foundation of China
under grant Nos.\ 11773052, 11333003; and 11761131016; and by a
China-Chile joint grant from Chinese Academy of Sciences (CAS) South
America Center for Astronomy (CASSACA). J.S.\ acknowledges support
from a Newton Advanced Fellowship awarded by the Royal Society
and the Newton Fund. M.B.\ acknowledges support from a Chinese Academy
of Sciences President's International Fellowship Initiative for
Visiting Scientists during part of this work, as well as the
hospitality of Shanghai Astronomical Observatory. This work made use
of the facilities of the Center for High Performance Computing at
Shanghai Astronomical Observatory.




\bibliography{test}


\end{document}